\documentclass[preprint]{aastex}

\newcommand\be{\begin{equation}}
\newcommand\ee{\end{equation}}

\input psfig
\shortauthors{Bahcall et al.}
\shorttitle{Solar models: current epoch, time dependences, neutrinos,
sound speeds}
\begin{document}
\doublespace

\title{
Solar Models: current epoch and time dependences, neutrinos,
and helioseismological properties
}

\author{John N. Bahcall}
\affil{Institute for Advanced Study, Olden Lane, Princeton, NJ 08540, U. S.
A.}
\author{M. H. Pinsonneault}
\affil{Department of Astronomy, Ohio State University, Columbus, Ohio
43210, U. S. A.}
\author{Sarbani Basu}
\affil{Astronomy Department, Yale University, P.O. Box 208101, New
Haven, CT 06520-8101, U.S.A.}

\begin{abstract}
We calculate accurate solar models and report the detailed time
dependences of important solar quantities.  We use helioseismology to
constrain the luminosity evolution of the sun and report the discovery
of semi-convection in evolved solar models that include diffusion.  In
addition, we compare the computed sound speeds with the results of
$p$-mode observations by BiSON, GOLF, GONG, LOWL, and MDI instruments.
We contrast the neutrino predictions from a set of eight standard-like
solar models and four deviant (or deficient) solar models with the
results of solar neutrino experiments.  For solar neutrino and for
helioseismological applications, we present present-epoch numerical
tabulations of characteristics of the standard solar model as a
function of solar radius, including the principal physical and
composition variables, sound speeds,  neutrino fluxes, and functions
needed for calculating solar neutrino oscillations.

\end{abstract}


\section{Introduction}
 \label{sec:introduction} Why are new calculations of standard solar
models of interest? After all, solar models have been used to
calculate neutrino fluxes since 1962 (Bahcall, Fowler, Iben, \& Sears
1963)\nocite{bahcall63} and solar atmospheres have been used to
calculate p-mode oscillation frequencies since 1970 (Ulrich
1970;\nocite{ulrich70} Leibacher \& Stein 1971).\nocite{leibacher71}
Over the past four decades, the accuracy with which solar models are
calculated has been steadily refined as the result of increased
observational and experimental information about the input parameters
(such as nuclear reaction rates and the surface of abundances of
different elements), more accurate calculations of constituent
quantities (such as radiative opacity and equation of state), the
inclusion of new physical effects (such as element diffusion), and the
development of faster computers and more precise stellar evolution
codes.

Solar models nevertheless remain at the frontiers of two different
scientific disciplines, solar neutrino studies and helioseismology. In
an era in which many major laboratory studies are underway to study
neutrino oscillations with the aid of very long baselines, $\sim 10^3$
km, between accelerator and detector, solar neutrinos have a natural
advantage, with a baseline of $10^8$ km (Pontecorvo
1968).\nocite{pontecorvo68} In addition, solar neutrinos provide
unique opportunities for studying the effects of matter upon neutrino
propagation, the so-called MSW effect (Wolfenstein 1978; Mikheyev \&
Smirnov 1985),\nocite{wolf,MiSm} since on their way to terrestrial
detectors they pass through large amounts of matter in the sun and, at
night, also in the earth . 

The connection with ongoing solar neutrino research imposes special
requirements on authors carrying out the most detailed solar modeling.
Precision comparisons between neutrino measurements and solar
predictions are used by many physicists to refine the determination of
neutrino parameters and to test different models of neutrino
propagation. Since the neutrino experiments and the associated
analysis of solar neutrino data are refined at frequent intervals, it
is appropriate to reevaluate and refine the solar model predictions as
improvements are made in the model input parameters, calculational
techniques, and descriptions of the microscopic and macroscopic
physics.

In this paper, we provide new information about the total solar
neutrino fluxes and the predicted neutrino event rates for a set of
standard and non-standard solar models. Using the best-available
standard solar model, we also present the calculated radial dependence
of the production rate for each of the important solar neutrino
fluxes.  We publish for the first time the results of a precision
calculation with the standard solar model of the electron density
throughout the sun, from the innermost regions of the solar core
to the solar atmosphere.  We also present for the first time a
detailed calculation of the radial profile of the number density of
scatterers of sterile neutrinos. These quantities are important for precision
studies of neutrino oscillations using solar neutrinos.

We also provide detailed predictions for the time evolution of some of
the important solar characteristics such as the depth and mass of the
solar convective zone, the radius and the luminosity of the sun, the central
temperature, density, pressure, and hydrogen mass fraction, as well as
the temperature, density,  pressure, and radiative opacity at the base
of the convective zone.  As far as we know, these are the first detailed
results submitted for publication on the time evolution of many of
these quantities. Some of the calculated time dependences  may be subject to
observational tests.

At the present writing, the sun remains the only main-sequence star
for which p-mode oscillations have been robustly detected. Thus only
for the sun can one measure precisely tens of thousands of the
eigenfrequencies for stellar pressure oscillations. The comparison
between the sound speeds and pressures derived from the observed
p-mode frequencies and those calculated with standard solar models has
provided a host of accurate measurements of the interior of the
nearest star.  The solar quantities determined by helioseismology
include the sound velocity
and density as a function of solar radius, the depth of the convective
zone, the interior rotation rate, and the surface helium abundance.
The excellent agreement between the
helioseismological observations and the solar model calculations
has shown that the large discrepancies
between solar neutrino measurements and solar model
calculations cannot be due to
errors in the solar models (cf. Figure~\ref{fig:diffbp00best}).

In this paper, we present a refined comparison between our best
standard solar model and measurements of the solar sound speeds
obtained using oscillation data from a number of different sources. We
use a
combination of data from the LOWL instrument and the BiSON network,
two sets of data from the GOLF instrument, as well as data from the
GONG network and the MDI instrument.

We describe in the following paragraph the organization of this
paper. But, since this paper contains a lot of information on
disparate topics, we recommend that the reader first turn to the final
section, \S~\ref{sec:discussion}, and peruse our summary and
discussion of the main new results. Section~\ref{sec:discussion} may
help the reader decide which sections of the paper the reader wants to
read (or to skip).  The different sections are written so that they
can be read more or less independently.

Section~\ref{sec:standard} defines the slightly-improved
Standard solar model and
describes the numerical tables that present details of the
contemporary characteristics of the Standard model.
Section~\ref{sec:timedependences} describes the main sequence
time dependences of
some of the most important characteristics of the Standard model. We
present in this section the time evolution of the solar radius and
luminosity, the properties of the convective zone, and the physical
characteristics of the center of the sun. We discuss solar properties
from the initial age main sequence to an age of 8 billion years.
Section~\ref{sec:variantdeviant} defines and compares the physical
characteristics of seven variant standard models and four deviant
(deficient) solar models, which together with the Standard model make
up a set of twelve models whose neutrino fluxes we evaluate in this
paper.  We previously used a subset of nine of these models to test the
robustness of helioseismological inversions (Basu, Pinsonneault, \&
Bahcall 2000, hereafter BPB2000).\nocite{basu2000} We discuss in this 
same section two new standard-like
solar models with heavy element to hydrogen
ratios that differ slightly from our previously adopted value of
$Z/X$. 
Section~\ref{sec:neutrinos} discusses solar neutrino
physics.  We present the predicted present-day neutrino fluxes for the
Standard model and for all of the variant and deviant solar models, as
well as the electron number density versus solar radius.  We contrast
the predicted neutrino event rates with the results of the chlorine,
Kamiokande, GALLEX, SAGE, Super-Kamiokande, and GNO solar neutrino
experiments. We also give in this section the calculated time
evolution of the most important solar neutrino fluxes.  Section~\ref
{sec:soundspeeds} compares calculated and observed sound speeds. We
present the results both on a panoramic scale suitable for discussing
the implications for solar neutrino physics and on a zoomed-in scale
appropriate for detailed investigations of solar physics.  In
\S~\ref{sec:hepluscz}, we compare, for all 12 of the solar models
discussed in this paper, the calculated values of the surface helium
abundance and the depth of the convective zone with the measured
values for these quantities.  We summarize and discuss our main
results in \S~\ref{sec:discussion}.

The interested reader may wish to consult the following works that
summarize the solar neutrino aspects of solar models (Bahcall
1989;\nocite{bahcall89} Bahcall \& Pinsonneault 1992,
1995;\nocite{bp92,bp95} Berezinsky, Fiorentini, \& Lissia
1996;\nocite{berezin96} Castellani et al. 1997;\nocite{cast97}
Richards et al. 1996;\nocite{ri96} Turck-Chi\`eze et
al. 1993;\nocite{turck93} Bahcall, Basu, \& Pinsonneault 1998\nocite{ba98})
and the helioseismologic aspects of solar models (Bahcall \& Ulrich
1988;\nocite{bu88} Bahcall \& Pinsonneault 1995;\nocite{bp95}
Christensen-Dalsgaard et al. 1996;\nocite{jcd96}; Guenther \& Demarque
1997;\nocite{guen97} Guzik 1998;\nocite{guzik98} Turck-Chi\`eze et
al. 1998;\nocite{turck98} Brun, Turck-Chi\`eze, \& Zahn
1999;\nocite{brun99} Fiorentini \& Ricci 2000).\nocite{fior00}

\section{Standard solar model}
\label{sec:standard}
We define in \S~\ref{subsec:definitionstandard} the Standard solar
model and present in \S~\ref{subsec:standardcharacteristics} some
of the important contemporary characteristics of the Standard solar
model, including detailed tables of the physical variables as a
function of the solar radius.

By `the Standard solar model', we mean the solar model which is
constructed with the best-available physics and input data. 
All of the solar models we consider, standard or `deviant' models,
(see below) are  required to fit the
observed luminosity and radius of the sun at the present epoch, as
well as the observed heavy element to hydrogen ratio at the surface of
the sun. No helioseismological constraints are used in defining the
Standard solar model.

Naturally, Standard models improve with time, as the input data are
made more accurate, the calculational techniques become faster and
more precise, and the physical description is more detailed.  Thus it
is necessary at each presentation of a new Standard model to describe
what has changed from the previous best model and to provide
references to the earlier work. The reader can see Bahcall
(1989)\nocite{bahcall89} for a general reference to the early work on
precise solar models that were motivated by the need to predict
accurate solar neutrino fluxes and to make comparisons with
helioseismological data.

\subsection{Definition of the Standard model}
\label{subsec:definitionstandard}

Our standard solar model\footnote{To simplify the language of the
discussion, we will often describe characteristics of the Standard
model as if we knew they were characteristics of the sun. We will
sometimes abbreviate the reference to this Standard model as BP2000 or
Bahcall-Pinsonneault 2000.} is constructed with the OPAL equation of
state (Rogers, Swenson, \& Iglesias 1996)\nocite{rog96} and OPAL
opacities (Iglesias \& Rogers 1996),\nocite{igl96} which are
supplemented by the low temperature opacities of Alexander \& Ferguson
(1994).\nocite{al94} The model was calculated using the usual mixing
length formalism to determine the convective flux.

The principal change in the input data is the use of the Grevesse \&
Sauval (1998)\nocite{gre98} improved standard solar composition in the
OPAL opacities (see http://www-phys.llnl.gov/Research/OPAL/index.htm)
and in the calculation of the nuclear reaction rates.  The refinements
in this composition redetermination come from many different sources,
including the use of more accurate atomic transition probabilities in
interpreting solar spectra.  The OPAL equation of state and the
Alexander and Ferguson opacities are not yet available with the
composition recommended by Grevesse and Sauval 1998.

We have used a present-epoch solar luminosity of $1369 {\rm~ W \, m^2}
[4 \pi \times {\rm (A.U.)}^2]$ for all the models described in detail in
this paper. Only for the neutrino fluxes presented in
\S~\ref{subsubsec:standardneutrino} and 
\S~\ref{subsubsec:nacre} have we used the more recent
best-estimate value of solar luminosity, $1366.2 {\rm ~W \,m^2} [4 \pi
\times ({\rm A.U.})^2] ~=~ 3.842 \times 10^{33} {\rm ~erg s^{-1}}$
(see Fr\"ohlich \& Lean 1998\nocite{frohlich98}; Crommelynck, Fichot,
Domingo, \& Lee 1996\nocite{crommel96}). The difference between
these two values for the luminosity is $0.2$\%. For the calculations
of uncertainties in neutrino flux predictions, we assume a $1\sigma$
uncertainty of $0.4$\%.  The uncertainty in the predicted solar
neutrino fluxes due to the luminosity is an order of magnitude smaller
than the total uncertainty in predicting the neutrino fluxes. For all
the other quantities we calculate, the uncertainty in the solar
luminosity has an even smaller effect.

The nuclear reaction rates were evaluated with the subroutine
exportenergy.f (cf. Bahcall \& Pinsonneault 1992),\nocite{bp92} using
the reaction data in Adelberger et al. (1998)\nocite{adel98} and with
electron and ion weak screening as indicated by recent calculations of
Gruzinov \& Bahcall (1998);\nocite{gruz98} see also Salpeter
(1954)\nocite{sal54}\footnote{Other approximations to screening are
sometimes used. The numerical procedures of Dzitko et
al. (1995)\nocite{dzitko95} and Mitler (1977)\nocite{mitler77} predict
reaction rates that are too slow for heavy ions because they assumed
that the electron charge density near a screened nucleus is the
unperturbed value, $en_e(\infty)$.  This assumption seriously
underestimates the charge density near heavy ions.  For example, it is
known that a screened beryllium nucleus under solar interior
conditions has charge density near the nucleus $\approx
-3.85en_e(\infty)$ (Gruzinov \& Bahcall 1997;\nocite{gruz97} Brown \&
Sawyer 1997;\nocite{brown97} all quantum mechanical calculations give
similar results, see Bahcall 1962,\nocite{bahcall62} and Iben, Kalata,
\& Schwartz 1967).\nocite{ibenks67}}.  The model incorporates helium
and heavy element diffusion using the exportable diffusion subroutine
of Thoul (cf. Thoul, Bahcall \& Loeb, 1994;\nocite{thoul4} Bahcall \&
Pinsonneault 1995)\nocite{bp95}\footnote{Both the nuclear energy
generation subroutine, exportenergy.f, and the diffusion subroutine,
diffusion.f, are available at the Web site www.sns.ias.edu/$\sim$jnb,
menu item: neutrino software and data.}.  
An independent and detailed treatment of diffusion by Turcotte et al.
(1998b)\nocite{turcotte98b} yields results for the impact of diffusion
on the computed solar quantities that are very similar to those
obtained here.  We have used the most recent and detailed calculation
(Marcucci et al. 2000a)\nocite{marcucci00a} for the $S_0$-factor for
the $^3$He(p,$e^+$~+~$\nu_e$)$^4$He reaction: $S_0({\rm hep}) = 10.1
\times 10^{-20}$ keV b, which is a factor of $4.4$ times larger than
the previous best-estimate [see \S~\ref{subsubsec:uncertainties},
Bahcall \& Krastev (1998),\nocite{bahcall98} and Marcucci et
al. (2000b)\nocite{marcucci00b} for a discussion of the large
uncertainties in calculating $S_0({\rm hep})$]. For values of
$S_0({\rm hep})$ in the range of current estimates, the assumed rate
of the $hep$ reaction only affects in a noticeable way the calculated
flux of $hep$ neutrinos and does not affect the calculated fluxes of
other neutrinos, the helioseismological characteristics, or other
physical properties of the sun.

For the standard model, the evolutionary calculations were started at
the main-sequence stage.  The model has a radius of 695.98 Mm.  We do
not study the pre-main sequence evolution in our standard model.  This
epoch has been discussed in the classical work of Iben
(1965)\nocite{iben65} and the effects of pre-main sequence have been
shown by Bahcall \& Glasner (1994)\nocite{baglas94} to be unimportant
for solar neutrino calculations (see also the thorough discussion by
Morel, Provost, \& Berthomieu 2000).\nocite{morel00} We do consider
one pre-main sequence model, which differs very little from the
corresponding model started at the zero-age main sequence.

 The ratio of heavy elements to hydrogen ($Z/X$) at the surface of the
model is 0.0230, which was chosen to be consistent with the value
obtained by Grevesse \& Sauval (1998).\nocite{gre98} A Krishna-Swamy
$T$-$\tau$ relationship for the atmosphere was used. We adopt a solar
luminosity $L_\odot = 3.844 \times 10^{33}~{\rm erg~s^{-1}}$ and a
solar age of $4.57 \times 10^9$~yr (see Bahcall \& Pinsonneault
1995).\nocite{bp95}

In the course of our recent analysis of systematic uncertainties in
helioseismological inversions (Basu, Pinsonneault, \& Bahcall
2000),\nocite{basu2000} we uncovered an error in the code we wrote for the
interpolation of the OPAL 95 opacities.  The edges of opacity tables
were flagged by zero opacity values; unfortunately, there were some
interpolation problems associated with the occurrence of zero values
inside the table. This problem occurred because the logarithm of the
opacity, which is what we were tabulating, can actually be zero.  The
interpolation error caused small changes in the opacity which produced
errors in the sound speed of order $0.1$\% for solar radii in the
range of $0.3 R\odot$ to $0.7 R_\odot$ and errors of order $3$\% in
the $^8$B neutrino flux (with smaller errors for other neutrino
fluxes).  In the following sections, we will point out more
specifically the changes in the neutrino predictions and in the sound
velocities that are produced by correcting this interpolation error.
These changes are particularly apparent in the comparison of
Figure~\ref{fig:differences6} and Figure~\ref{fig:differences5}, which
are discussed in the helioseismology section,
\S~\ref{sec:soundspeeds}.

\subsection{Some contemporary characteristics of the standard solar model}
\label{subsec:standardcharacteristics}

The details of the structure of a standard solar model are of
importance for both helioseismology and solar neutrino
calculations. In the first paper in this series in which we published
the details of a model of the solar interior structure (Bahcall et
al. 1982),\nocite{bahcall82} we presented a table with only $27$
radial shells (rows of numbers) and $10$ variables for each shell
(mass, radius, temperature, density, hydrogen fraction, helium
fraction, luminosity generated, and the source density of the $p-p$,
$^7$Be, and $^8$B neutrinos). Over the years, much greater precision
was required by the increasing sophistication of the
helioseismological measurements and the solar neutrino
calculations. Fortunately, the computing capacity more than made up
for the necessary increase in model details.

We have created from the output of the present calculations two
exportable computer files that contain in easily readable form the
details of our best Standard solar model (BP2000).  Physical variables
are given at $875$ separate radial shells, which is the same number of
shells used to calculate the solar interior model. In addition to the
variables cited above, this file contains the pressure, electron
number density, the mass fractions of $^3$He, $^7$Be, $^{12}$C,
$^{14}$N, and $^{16}$O, as well as the source densities of all eight
of the most important solar neutrino fluxes. These files are
accessible at http://www.sns.ias.edu/$\sim$jnb. Previous standard
solar models in this series (published in 1982, 1988, 1992, 1995, as
well as 1998) are available at the same URL and can be used to test
the robustness of conclusions that depend upon models of the solar
interior.

\section{Time dependences}
\label{sec:timedependences}
In this section, we present and discuss the time dependences of some
of the principal characteristics of the standard solar model. We begin
in \S~\ref{subsec:radiusetal} by describing the separate temporal
evolution of the solar radius and the solar luminosity and then
discuss the simple relation we have found between the values of these
two quantities as a function of time.  We also demonstrate that the
solar luminosity is a robust function of time. In
\S~\ref{subsec:energyfractions}, we present the time-dependent
fractions of the solar luminosity that are produced by different
nuclear fusion reactions. We concentrate in \S~\ref{subsec:convective}
on the convective zone, presenting the calculated time dependences of
the depth and the mass of the convective zone. We also report the time
dependence of the temperature, density, pressure, and opacity at the
base of the convective zone. In \S~\ref{subsec:centralvalues}, we
discuss the time dependence of quantities at the center of the sun,
the central temperature, density, pressure, and hydrogen mass
fraction. In \S~\ref{subsec:separations}, we calculate and discuss the
large and small separations of the $p$-mode frequencies as a function
of solar age.  Since we do not discuss pre-main sequence evolution,
our calculations are not precise for times less than $0.1 \times
10^9$~yr (see Morel, Provost, \& Berthomieu 2000\nocite{morel00};
Weiss \& Schlattl 1998).\nocite{weiss98}

Iben (1967,1974)\nocite{iben67,iben74} and Demarque \& Guenther
(1991)\nocite{demarque91} summarize in comprehensive presentations the
evolution of solar parameters in models that were calculated prior to
the inclusion of element diffusion in solar evolutionary codes.  These
discussions did not encounter the problem of semi-convection discussed
here in \S~\ref{subsec:convective} because this phenomenon is
caused by effects of diffusion near the base of the convective zone.

The solar radius and luminosity (or equivalently, the solar effective
temperature and luminosity) constitute precise constraints on the
possible geological histories of the earth. We quantify these
constraints  in the following subsection and specify upper limits to
the allowed discrepancies from the standard solar model profile of
solar luminosity versus age.

\begin{figure}[!ht]
\centerline{\psfig{figure=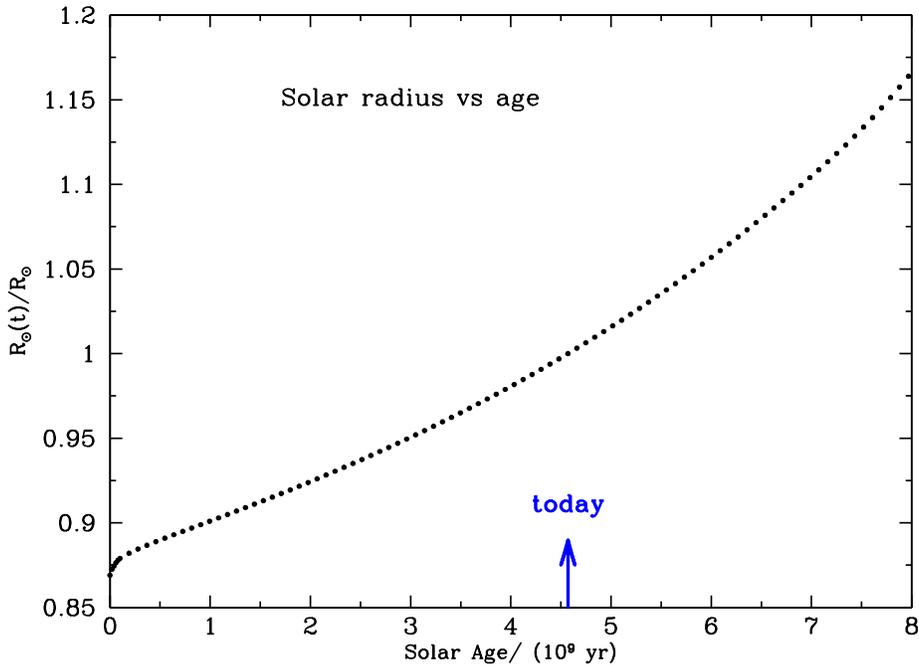,width=5in,angle=270}}
\caption[]{\baselineskip=12pt The calculated radius, $R_\odot(t)$, as a
function of age for the standard solar model, Bahcall-Pinsonneault
(2000). The solar age is
measured in units of $10^9$ yr. The present age of the sun,
$4.57\times 10^9$ years, is indicated by an arrow in
Figure~\ref{fig:rvsageplus}.
The radius increases from $0.87 R_\odot$ at the zero age main sequence
to $1.0 R_\odot$ at the present epoch and $1.18 R_\odot$ at a solar age
of $8$ billion years.
\label{fig:rvsageplus}}
\end{figure}

\begin{table}[!htb]
\centering
\singlespace
\caption[]{\baselineskip=12pt
The solar radius, $R_\odot (t)$, as a function of solar age for the
Standard solar model. The present age of the sun is $4.57 \times 10^9$
years (see Bahcall \& Pinsonneault 1995)\nocite{bp95}\label{tab:radiusvstime}}
\begin{tabular}{ccccc}
\tableline\tableline
Age&$R_\odot (t)$&&Age&$R_\odot (t)$\\
($10^9$ yr)&($R_\odot$ (today))&&($10^9$ yr)&($R_\odot$ (today))\\
\tableline
0.0&    0.869&&   4.2&    0.987\\
0.2&    0.882&&   4.4&    0.994\\
0.4&    0.888&&   4.6&    1.001\\
0.6&    0.892&&   4.8&    1.008\\
0.8&    0.897&&   5.0&    1.016\\
1.0&    0.901&&   5.2&    1.023\\
1.2&    0.906&&   5.4&    1.031\\
1.4&    0.910&&   5.6&    1.040\\
1.6&    0.915&&   5.8&    1.048\\
1.8&    0.920&&   6.0&    1.057\\
2.0&    0.924&&   6.2&    1.066\\
2.2&    0.929&&   6.4&    1.075\\
2.4&    0.934&&   6.6&    1.085\\
2.6&    0.940&&   6.8&    1.095\\
2.8&    0.945&&   7.0&    1.105\\
3.0&    0.951&&   7.2&    1.116\\
3.2&    0.956&&   7.4&    1.127\\
3.4&    0.962&&   7.6&    1.139\\
3.6&    0.968&&   7.8&    1.152\\
3.8&    0.974&&   8.0&    1.166\\
4.0&    0.981\\
\tableline
\end{tabular}
\end{table}

\subsection{Radius and luminosity versus age}
\label{subsec:radiusetal}

Figure~\ref{fig:rvsageplus} and Table~\ref{tab:radiusvstime} present
the calculated radius of the sun versus solar age for the standard
solar model. The values given are normalized to the present-day solar
radius, $R_\odot$. Over the lifetime of the sun, the model radius has
increased monotonically from an initial value of $0.869 R_\odot$ to
the current value $1.0 R_\odot$, a $15$\% rise. At a solar age of $8$ billion
years, the solar radius will increase to $1.17 R_\odot$.  We shall see
in the following discussion that some of the important evolutionary
characteristics of the sun can be expressed simply in terms of the
ratio $R_\odot(t)/R_\odot({\rm today})$.

\begin{figure}[!ht]
\centerline{\psfig{figure=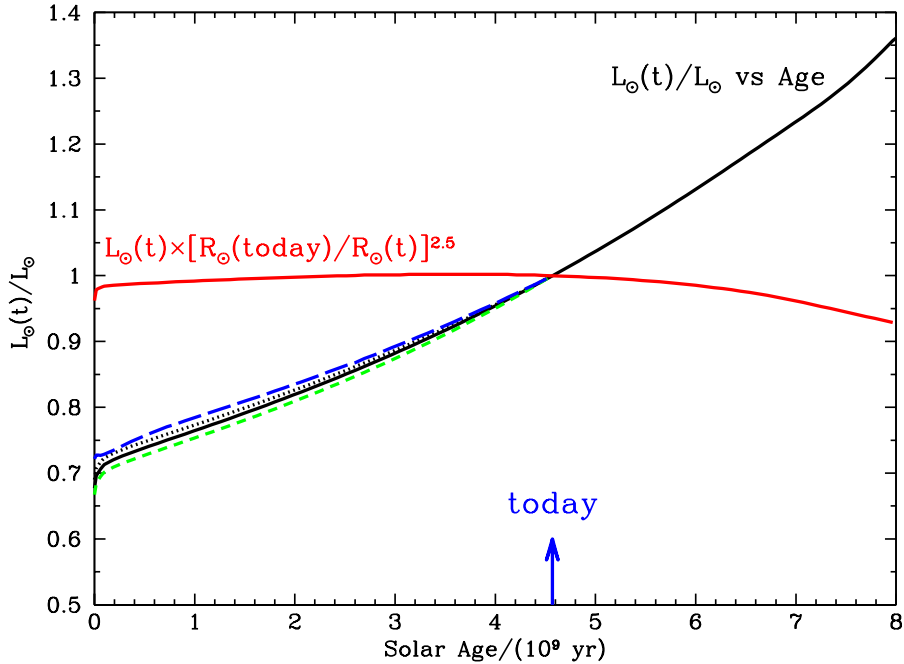,width=5in,angle=270}}
\caption{\baselineskip=12pt The normalized solar luminosity,
$L_\odot(t)/L_\odot({\rm today})$ versus solar age for the Standard
solar model (solid curve) and for three `deficient' solar models: the
No Diffusion model (dotted curve), the $S_{34} = 0$ model (short
dashes), and the Mixed model (long dashes). The luminosity evolution
of the sun is essentially the same in all solar models we have
investigated, including deficient solar models.  The rms deviation of
the deviant models from the standard solar model luminosity is only
$1$\% over the history of the sun from the zero-age main sequence to
the current epoch (see text for more details).  The product
$L_\odot(t) \times R_\odot(t)^{-2.5}$ varies by $\pm 4$\% over the
entire period from the zero age main sequence to a solar age of $8$
billion years, while the solar luminosity itself varies by slightly
more than a factor of two during this period. In the period between 4
billion years to 8 billion years, the relation $L_\odot(t) \propto
R_\odot(t)^2$ is satisfied to $\pm 1/2\%$. The solar luminosity has
increased by $48$\% from the zero main sequence to the present
epoch. The present age of the sun is indicated by an arrow at $4.59
\times 10^9$ years.\label{fig:lumnormalized} }
\end{figure}

\begin{table}[!htb]
\centering
\singlespace
\caption[]{\baselineskip=12pt
Solar luminosity as a function of solar age for the
Standard solar model.  The table gives the computed values of the
solar luminosity in units of the present-day solar luminosity.  The
model was iterated to give the observed luminosity at the present
epoch, $4.57 \times 10^9$ yr.\label{tab:lumstandard}}
\begin{tabular}{ccccc}
\tableline\tableline
Age&$L_\odot (t)$&&Age&$L_\odot (t)$\\
($10^9$ yr)&($L_\odot$ (today))&&($10^9$ yr)&($L_\odot$ (today))\\
\tableline
0.0&    0.677&&4.2&    0.970\\
0.2&    0.721&&4.4&    0.986\\
0.4&    0.733&&4.6&    1.003\\
0.6&    0.744&&4.8&    1.020\\
0.8&    0.754&&5.0&    1.037\\
1.0&    0.764&&5.2&    1.055\\
1.2&    0.775&&5.4&    1.073\\
1.4&    0.786&&5.6&    1.092\\
1.6&    0.797&&5.8&    1.112\\
1.8&    0.808&&6.0&    1.132\\
2.0&    0.820&&6.2&    1.152\\
2.2&    0.831&&6.4&    1.172\\
2.4&    0.844&&6.6&    1.193\\
2.6&    0.856&&6.8&    1.214\\
2.8&    0.869&&7.0&    1.235\\
3.0&    0.882&&7.2&    1.256\\
3.2&    0.896&&7.4&    1.278\\
3.4&    0.910&&7.6&    1.304\\
3.6&    0.924&&7.8&    1.332\\
3.8&    0.939&&8.0&    1.363\\
4.0&    0.954\\
\tableline
\end{tabular}
\end{table}

Figure~\ref{fig:lumnormalized} and Table~\ref{tab:lumstandard} show
the calculated solar luminosity as a function of solar age, normalized
to the present-day solar luminosity, $L_\odot$.  For the standard
model, the total solar luminosity has risen monotonically from a
zero-age value of $0.677 L_\odot$ .

The time evolution of the solar luminosity is robust. We also show in
Figure~\ref{fig:lumnormalized} the solar luminosity as a function of
time for the three most deficient solar models that are described in
the following section, \S~\ref{sec:variantdeviant}.  The rms
difference between the standard luminosity and the luminosity of the
deviant models is $1.6$\% for the mixed model ($1.2$\% ignoring the
first Gyr), $0.7$\% for the no diffusion model ($0.5$\% ignoring the
first Gyr), and $0.9$\% for the $S_{34} = 0$ model ($0.8$\% ignoring
the first Gyr).  The largest deviations occur for the zero-age main
sequence models and are $2.5$\% for the mixed model, $1.9$\% for the
no diffusion model, and $1.7$\% for the $S_{34} = 0$ model.  All of
the solar models show essentially the same shape for the luminosity
evolution as a function of age.

Figure~\ref{fig:lumnormalized}
shows that the product
\begin{equation}
L_\odot(t) \times R_\odot(t)^{-2.5}~=~ {\rm  constant}
\label{eqn:lrproportional}
\end{equation}
to an accuracy of about $\pm 4$\% over the entire period from the zero age
main-sequence to an age of $8$ billion years.
The solar luminosity $L_\odot(t)$ varies from $0.68 L_\odot$ to
$L_\odot$ at the present epoch and will reach $1.36 L_\odot$
after $8$ billion years on the main sequence. The
corresponding values of $L_\odot(t) \times
[R_\odot ({\rm today})/R_\odot(t)]^{2.5}$ are $0.96$, $1.00$, and
$0.93$.
Thus the luminosity of the sun varies by slightly more than a factor
of two while $L_\odot(t) \times
[R_\odot ({\rm today})/R_\odot(t)]^{2.5}$ varies by only a few percent.

Over the first 5 billion years, Eq.~(\ref{eqn:lrproportional}) is
satisfied to an accuracy of $\pm 2\%$.  From 4 billion years to 8
billion years, the relation is somewhat less steep, $L_\odot(t)
\propto R_\odot(t)^2$ to an accuracy of $\pm 1/2\%$.  This transition
from $L_\odot (t) \propto R_\odot(t)^{2.5}$ to $L_\odot(t) \propto
R_\odot (t)^2$ can be seen clearly in Figure~\ref{fig:lumnormalized}.

\begin{table}[!htb]
\centering
\singlespace
\caption[]{\baselineskip=12pt
The calculated effective temperature as a function of solar
age for the Standard model.  The solar age is measured in units of
$10^9$~yr and the solar effective temperature is measured in units of
the present-epoch effective temperature\label{tab:teff}}
\begin{tabular}{lcccc}
\tableline\tableline
\multicolumn{1}{c}{Age}&$T_{\rm eff}$&&\multicolumn{1}{c}{Age}&$T_{\rm
eff}$\\
\tableline
0.0& 0.973&& 4.5& 1.000\\
0.5& 0.983&& 5.0& 1.001\\
1.0& 0.985&& 5.5& 1.002\\
1.5& 0.987&& 6.0& 1.003\\
2.0& 0.990&& 6.5& 1.003\\
2.5& 0.992&& 7.0& 1.003\\
3.0& 0.994&& 7.5& 1.001\\
3.5& 0.996&& 8.0& 1.000\\
4.0& 0.998\\
\tableline
\end{tabular}
\end{table}

Table~\ref{tab:teff} gives the calculated effective temperature of the
Standard model,  $T _{\rm eff}$, as a function of solar age.
The effective temperature 
varies by only $\pm 1.5$\% over the entire period from the zero age
main sequence to an age of $8$ billion years and only by $\pm 0.7$\%
from $2$ billion years to $8$ billion years.

The relation between $L_\odot (t)$ and $R_\odot (t)$ discussed above
can be restated using the calculated time-dependence of the solar
effective temperature, which is summarized in Table~\ref{tab:teff}.  In
the period between $4$ Gyr and $8$ Gyr, the effective temperature is
essentially constant (to an accuracy of $\pm 0.25$\%).  This constancy
of the effective temperature results in the scaling relation
$L_\odot(t) \propto R_\odot(t)^2$ that is valid in this period.  The
effective temperature changes by an order of magnitude larger fraction
during the evolution up to the present-age sun, which results in a
dependence closer to $L_\odot \propto R_\odot^{2.5}$ (cf. also
Figure~2 of Demarque \& Guenther 1991).\nocite{demarque91}

\subsection{Energy fractions}
\label{subsec:energyfractions}

\begin{figure}[!htb]
\centerline{\psfig{figure=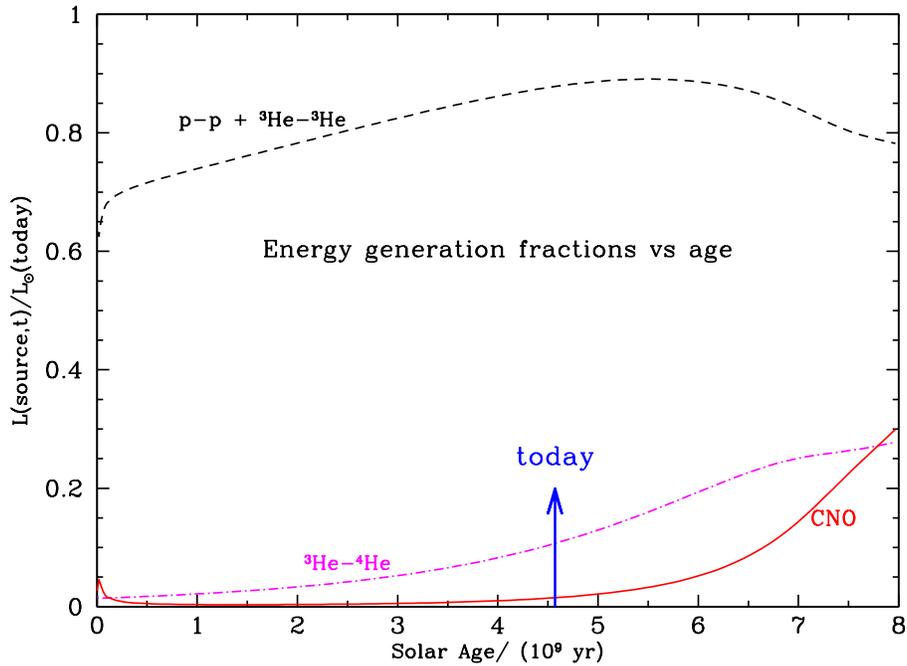,width=5in,angle=270}}
\caption{\baselineskip=12pt The fraction in the Standard model of the solar luminosity
produced by different nuclear fusion reactions versus solar age. The
luminosity generated by the $p-p$ nuclear fusion branch that is
terminated by the $^3$He-$^3$He reaction is marked by a dashed curve
in the figure and the luminosity produced by the $p-p$ branches that
proceed through the $^3$He-$^4$He reaction is denoted by a dot-dashed
curve. The luminosity generation by the CNO cycle is indicated by a
solid line. The unit of luminosity is the present-day total solar
luminosity. At the present epoch, the $p-p$ + $^3$He-$^3$He
reactions produce $87.8$\% of the solar luminosity and the branches
terminating through the $^3$He-$^4$He reaction generate $10.7$\% of
the solar luminosity. The CNO cycle produces $1.5$\% of the
present-epoch luminosity.\label{fig:efractions}}
\end{figure}

Figure~\ref{fig:efractions} shows, for the Standard model, the energy
generated by different nuclear fusion reactions as a function of solar
age. The present-day total solar luminosity, $L_\odot({\rm today})$,
is the unit of luminosity in Figure~\ref{fig:efractions}.

The branch of the $p-p$ chain that is denoted in
Figure~\ref{fig:efractions} by $p-p$ + $^3$He-$^3$He (the dashed
curve) proceeds primarily through the reactions $p(p,e^+~\nu_e)^2{\rm
H}(p,\gamma){\rm^3He}({\rm ^3He},2p){\rm^4He}$.  For simplicity, we
include all $p-p$ reactions in this sum but do not show explicitly the
$pep$ reactions in the above scheme. The small energy contribution due
to $pep$ reactions is included in the calculations that led to
Figure~\ref{fig:efractions}. The $^3$He-$^4$He branch (the dot-dashed
curve) includes the nuclear reactions that produce both the $^7$Be and
the $^8$B neutrinos: ${\rm ^3He}({\rm^4He},\gamma){\rm
^7Be}(e^-,\nu_e){\rm^7Li}(p,{\rm^4He}){\rm^4He}$ and
${\rm^3He}({\rm^4He},\gamma){\rm^7Be}(p,\gamma){\rm^8B}({\rm^4He} +
e^+ + \nu_e){\rm^4He}$.  The CNO reactions are denoted by CNO in
Figure~\ref{fig:efractions}.

The branch that terminates via the $^3$He-$^3$He reaction dominates
the solar energy generation throughout the main sequence lifetime
shown in Figure~\ref{fig:efractions}, although the CNO reactions
overtake the $^3$He-$^4$He branches at an age of about eight billion
years. At an age of one billion years, $96.7$\% of the solar
luminosity is generated through $p-p$ reactions plus the $^3$He-$^3$He
termination, $2.8$\% through the $^3$He-$^4$He termination, and only
$0.5$\% through the CNO cycle. The situation gradually changes as the
sun heats up. At the present epoch, the $^3$He-$^3$He termination
produces $87.8$\% of the solar luminosity and the branches terminating
through the $^3$He-$^4$He reaction generate $10.7$\% of the solar
luminosity. The CNO cycle produces $1.5$\% of the present-epoch
luminosity. By the time the Standard solar model reaches an age of
eight billion years, the percentages are $57.6$\%, $20.4$\%, and
$22.0$\%, respectively, for $^3$He-$^3$He, $^3$He-$^4$He, and CNO
reactions. The energy loss due to gravitational energy expansion
ranges from $-0.03$\% to $-0.04$\% to $-0.07$\% as the sun's age
increases from one billion years to the present epoch to eight billion
years.

\subsection{Convective zone}
\label{subsec:convective}

\begin{figure}[!htb]
\centerline{\psfig{figure=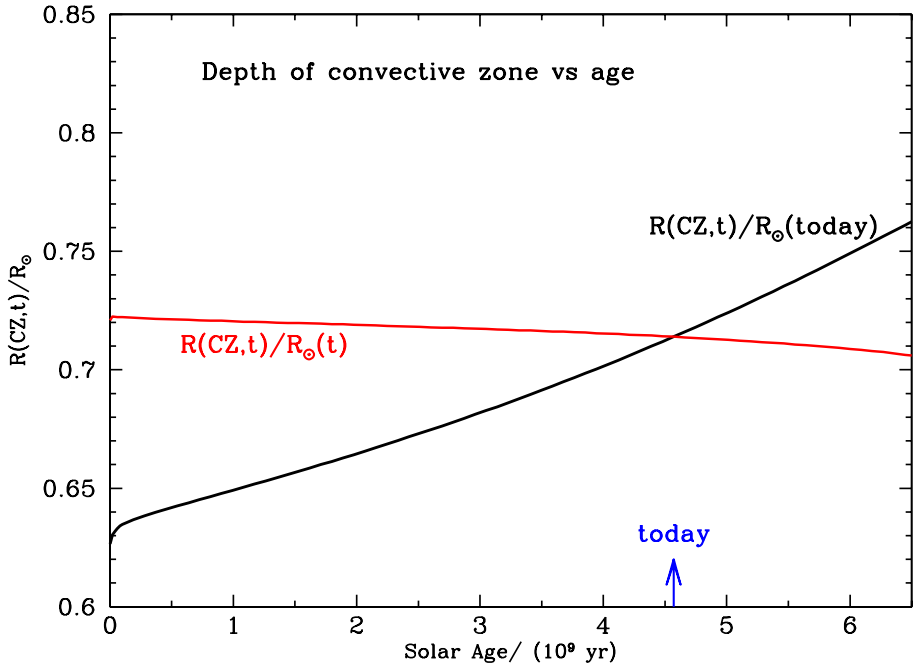,width=5in,angle=270}}
\caption[]{\baselineskip=12pt 
The depth of the convective zone, $R({\rm CZ},t)$, as a
function of age for the standard solar model.
The depth of the convective zone is approximately proportional to the
contemporary solar radius.
The solar age is
measured in units of $10^9$ yr.
\label{fig:czvsageplus}}
\end{figure}

\begin{figure}[!htb]
\centerline{\psfig{figure=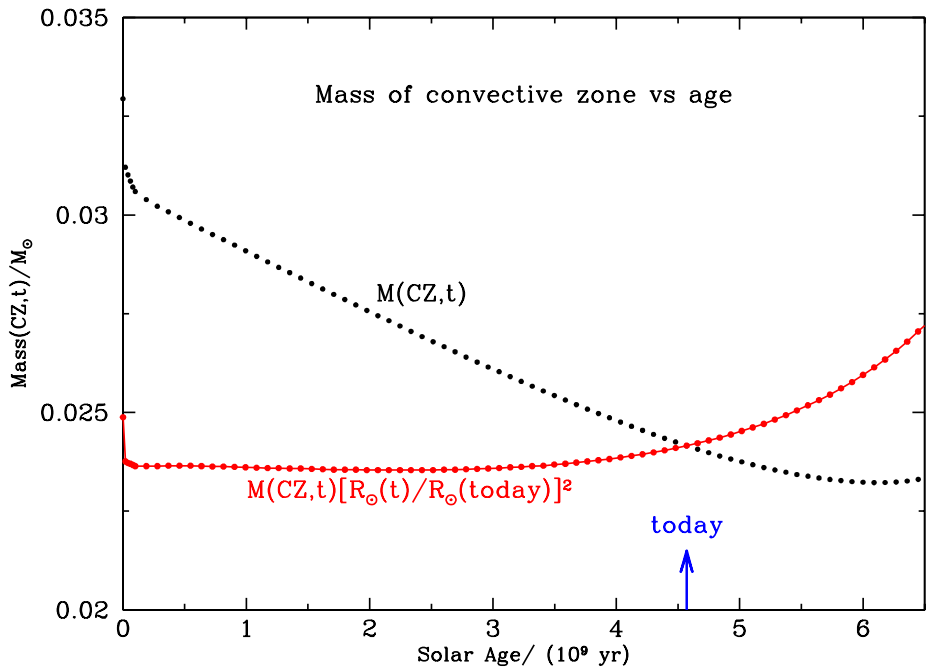,width=5in,angle=270}}
\caption[]{\baselineskip=12pt
The mass included within the convective zone, $M({\rm
CZ},t)$, as a function of age for the standard solar model. The mass
of the convective zone is measured in units of $M_\odot({\rm today})$
and the solar age is measured in units of $10^9$
yr.  The mass of the convective zone is approximately proportional to $R_\odot(t)^{-2}$.\label{fig:czmass}}
\end{figure}

Figure~\ref{fig:czvsageplus} shows the depth of the convective
zone,$R({\rm CZ},t)$, as a function of age for the standard solar
model.  Correspondingly, Figure~\ref{fig:czmass} shows the mass of the
convective zone, $M({\rm CZ},t)$, as a function of age. In both cases,
the temporal dependence from the zero age main sequence  to the current
epoch is describable by a simple function of $R_\odot(t)/R_\odot({\rm
today})$.

We limit the time period covered in Figure~\ref{fig:czvsageplus} and
Figure~\ref{fig:czmass} to be less than $6.5$ billion years, since
between $6.5$ and $7.0$ billion years of age semi-convection begins to
influence the computed $R({\rm CZ},t)$ and $M({\rm CZ},t)$. The
evolution of the depth of the solar convective zone was previously
studied by Demarque \& Guenther (1991)\nocite{demarque91} 
in an investigation that did
not include element diffusion.

The occurrence of semi-convection in evolved solar models is discussed
here for the first time, to the best of our knowledge. The onset of
semi-convection is triggered by the effects of element diffusion, which
was first included in detailed solar models in the early 1990's (see,
e.g., Bahcall \& Loeb 1990;\nocite{bahcall90} 
 Proffitt \& Michaud 1991;\nocite{proffitt91} Bahcall \&
Pinsonneault 1992).\nocite{bp92}

The process works essentially as follows.  The time scale for
diffusion decreases as the surface convective zone becomes shallower;
the metal abundance increases steadily below the surface convective
zone.  At some point in the evolutionary history, near an age of $6.5$
billion years, the opacity from the enhanced metal abundance below the
surface convective zone becomes large enough to make the metal-rich
radiative layers below the surface convective zone convectively
unstable.  However, the mixing of the metals into the envelope causes
a local drop in the metal abundance $Z$ and and the opacity $\kappa$
at the base of the convective zone. This result in turn causes the
convective zone to recede until the metal abundance builds up again.
Richer, Michaud, \& Turcotte (2000)\nocite{richer00} discuss a
related phenomenon in A and F stars.

We noticed the existence of semi-convection in our standard solar
model only because we made precise plots of the depth and mass of the
convective zone as a function of time (see Figure~\ref{fig:czvsageplus}
and Figure~\ref{fig:czmass}).  The effects of semi-convection were not
noticeable in plots of external quantities such as the solar
luminosity or effective temperature (see
\S~\ref{subsec:radiusetal} ).  In our current code, if a region
is convectively unstable according to the Schwarzschild criterion,
then we instantly mix this material with the other material in the
convective zone. A more accurate treatment, which allows for the
possibility of a composition gradient being established at the base of
the convective zone, is required to calculate reliably the influence
of semi-convection once it begins to be perceptible in the numerical
solutions [see, for example, Merryfield (1995),\nocite{merryfield95} 
Canuto (2000)\nocite{canuto00} and
references quoted therein]. This is why we have terminated the plots
of$R({\rm CZ},t)$ and $M({\rm CZ},t)$ at a solar age of $6.5 \times
10^9$ yr.

From the zero age main sequence to the present solar age, the depth of
the convective zone is approximately proportional to the contemporary
solar radius, $R_\odot(t)$, i.e.,

\begin{equation}
0.714 \leq R({\rm CZ},t)/R_\odot(t) \leq 0.722 .
\label{eqn:rczrproportional}
\end{equation}
The zero-age main sequence value is $R({\rm
CZ}, t\, = \, 0) = 0.627 R_\odot({\rm today})$ and the present-day
value is $R({\rm CZ, today}) = 0.714 R_\odot({\rm today})$, which
corresponds to an approximately $14$\% decrease in the depth of the
convective zone over the age of the sun. However, if we normalize the
depth of the convective zone to the contemporary solar radius, the
change is very small, $R({\rm CZ}, t\, = \, 0) = 0.722 R_\odot(t \, =
\, 0)$ and $R({\rm CZ},{\rm today}) = 0.714 R_\odot({\rm
today})$. After $6.5$ billion years, $R({\rm CZ}, t\, = \, 6.5) =
0.764 R_\odot({\rm today})$ and, in terms of the contemporary
solar radius, $R({\rm CZ}, t\, = \, 6.5) = 0.706 R_\odot(t\, = \,
6.5)$. The relative evolution between the convective zone depth and
the total solar radius, $ d [R({\rm CZ}, t)/R_\odot(t)]/d t = -0.002$
per $10^9$ yr, is very slow.

Over the period from the initial age main sequence to the present
epoch, the mass of the solar convective zone decreases with time as
$R_\odot(t)^{-2}$,

\begin{equation}
M({\rm CZ},t) \times R_\odot(t)^2 ~=~{\rm constant}
\label{eqn:mczrproportional}
\end{equation}
 to an accuracy of about $\pm 1$\%.  The zero-age main
sequence value of the mass included within the convective zone is
$0.0329 M_\odot$, which decreases to $0.02415 M_\odot$ at the present
epoch. However, this proportionality is no longer valid for larger
ages, as can be seen in Figure~\ref{fig:czmass}.  The mass of the
convective zone is $M({\rm CZ},6.5) = 0.0233 M_\odot$ at $6.5$ billion years.
At about $6.8$ billion years, the previous monotonic decrease of the
mass of the convective zone is reversed and $M({\rm CZ},t)$ begins to
increase with time. This behavior is not shown in
Figure~\ref{fig:czmass} since the calculated increase in the mass of the
convective zone occurs in the same time frame as the onset of
semi-convection.

\begin{figure}[!htb]
\centerline{\psfig{figure=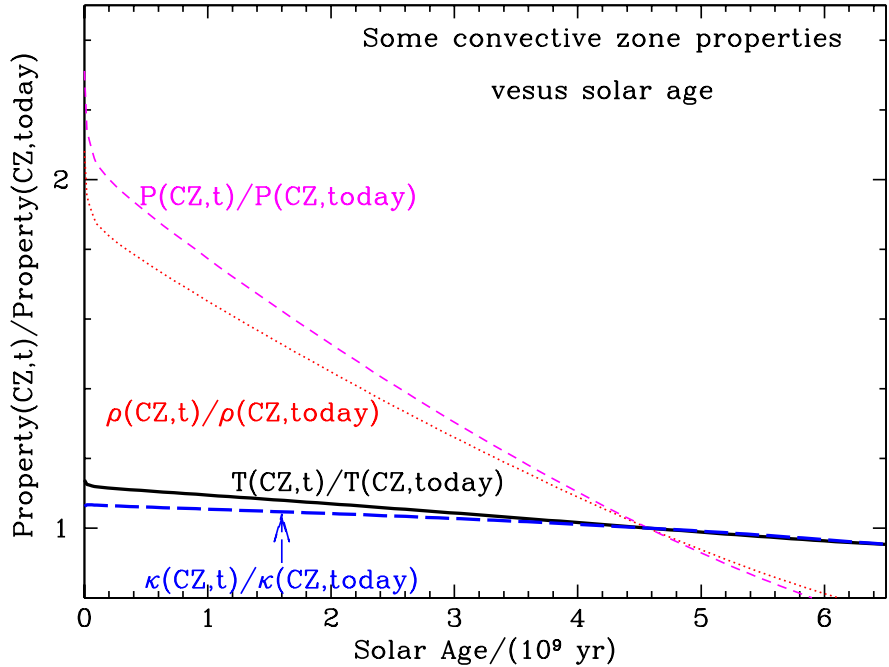,width=5in,angle=270}}
\caption[]{\baselineskip=12pt
Some properties at the base of the convective zone as a
function of age for the Standard solar model. The properties shown are
the temperature, T(CZ,t) (solid curve), the density, $\rho(CZ,t)$
(dotted), the pressure, $P(CZ,t)$ (short dashes), and the radiative
opacity, $\kappa(CZ,t)$ (long dashes). All of the quantities are
normalized by dividing by their values at the present epoch.  After
$6.5$ billion yr, semi-convection begins to be important.
\label{fig:cznormalizeprop}}
\end{figure}

Figure~\ref{fig:cznormalizeprop} shows some properties at the base of
the convective zone as a function of solar age. The figure displays
the time dependence of the temperature, T(CZ,t) (solid curve), the
density, $\rho({\rm CZ},t)$ (dotted), the pressure, $P({\rm CZ},t)$
(short dashes), and the radiative opacity, $\kappa({\rm CZ},t)$ (long
dashes). For convenience in plotting, each of the physical variables
has been divided by its value at the current epoch. In cgs units, the
Standard model parameters have the following
values at an age of $4.57\times 10^9$
yr: $T({\rm CZ,today}) = 2.18 \times 10^6$ K, $\rho({\rm CZ,today}) =
0.19$, $P({\rm CZ,today}) = 5.58 \times 10^{13}$, and $\kappa({\rm
CZ,today}) = 20.5$.

The temperature at the base of the convective zone decreases by $14$\%
from the zero-age main-sequence to the current solar age, i. e., to a
good approximation

\begin{equation}
T({\rm CZ},t) \propto R({\rm CZ},t)^{-1}.
\label{eqn:tproptoRminus1}
\end{equation}
The opacity at the base of the convective zone decreases by $6$\% over
the same period.  The density and the pressure at the base of the
convective zone decrease by much larger quantities, by factors of
$2.1$ and $2.3$, respectively.  Equation~\ref{eqn:tproptoRminus1} is
valid to an accuracy of about $\pm 1\%$
throughout the $6.5\times 10^9$ yr shown in
Figure~\ref{fig:cznormalizeprop}.

The base of the convective zone is defined by the Schwarzschild
criterion.  Because the adiabatic gradient is nearly constant in time,
the equations of stellar structure imply that the quantity $\kappa
PL/MT^4$ will be approximately constant at the base of the convective
zone.  From Figure~\ref{fig:cznormalizeprop}, we see that both the
opacity and the temperature decrease slowly at the base of the
convective zone.  Solar models therefore compensate for the increase
of the luminosity by the decrease of the pressure at the boundary
between radiative and convective equilibrium.

\subsection{Central values of Temperature, Density, and Pressure}
\label{subsec:centralvalues}

\begin{figure}[!ht]
\centerline{\psfig{figure=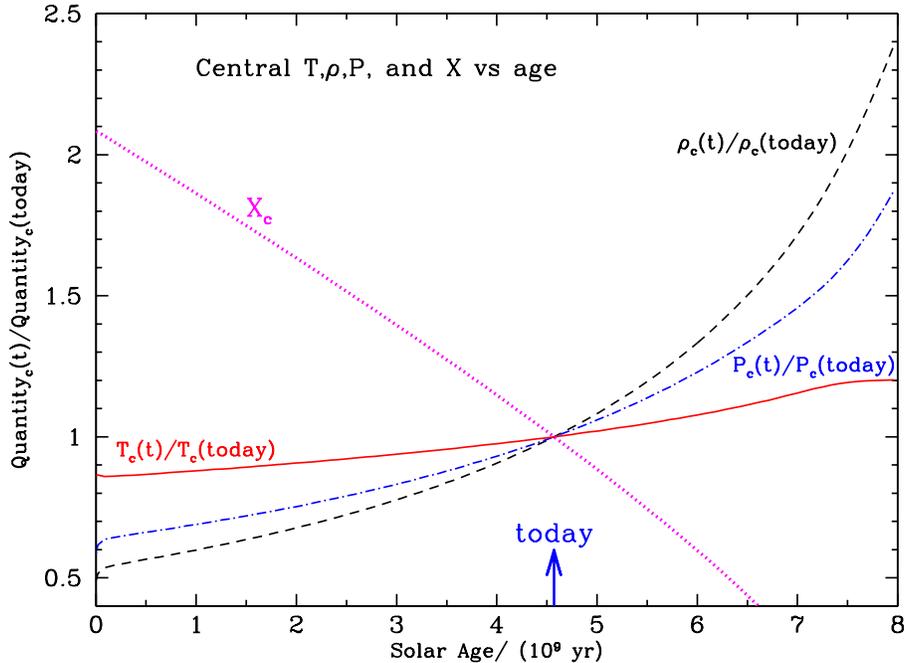,width=5in,angle=270}}
\caption{\baselineskip=12pt The temporal evolution of the central
temperature, density, pressure, and hydrogen mass fraction. The figure
shows the computed values for the Standard solar model of the central
temperature (solid line), pressure (dot-dash line), density (dash
line), and hydrogen mass fraction (dotted line).
\label{fig:trhop}}
\end{figure}

Figure~\ref{fig:trhop} shows the time dependence of the central values
for the temperature, density, and pressure of the Standard solar
model. The results are normalized to the computed values for the
present epoch.

Over the $8$ billion years shown in Figure~\ref{fig:trhop}, the
central temperature increases by about $39$\%.  At the zero-age main
sequence, the central temperature is $13.5 \times 10^9$K and is $15.7
\times 10^9$K at the current epoch; the central temperature of the
model reaches $18.8 \times 10^9$K at a solar age of $8$ billion years.
This increase is very similar to that of the solar radius. In fact,
the central value of the temperature is, to a reasonable
approximation, proportional to the solar radius, $R_\odot(t)$. The
ratio
\begin{equation}
T_c(t)/R_\odot(t) ~=~ {\rm constant}
\label{eqn:tcoverr}
\end{equation}
to an accuracy of $\pm 1.5$\% ($\pm 3$\%) from the zero ago-main
sequence to the current epoch (to a solar age of $6.5$ billion years).

The sun derives its luminosity by burning hydrogen into helium. The
hydrogen mass fraction at the center of the sun, $X_c$, decreases from
$0.708$ at the zero-age main sequence to $0.340$ at the current epoch,
a decrease by more than a factor of two. At an of $6.5$ billion years,
the mass fraction has decreased to $0.145$, a decrease by a factor of
almost five from the zero-age value. The remaining hydrogen is burned
very quickly, with $X_c$ reaching $0.008$ at a solar age of $8$
billion years.

\subsection{Large and small $p$-mode separations}
\label{subsec:separations}

The $p$-mode oscillation frequencies change as the structure of the
sun evolves with age.  Since only low degree modes (small spherical
harmonic, $l$) are observable for stars other than the Sun, we
shall discuss the changes in just these modes.

For the high-order (large radial node, $n$), low-degree modes
that are expected to be observed in stars,  the
frequencies satisfy the following relation to a good approximation
(e.g., Tassoul 1980)\nocite{tassoul80}:
\begin{equation}
\nu_{nl} \simeq \Delta \left( n + {l \over 2} + \alpha \right ) .
\label{eq:largesep}
\end{equation}
Here $\nu_{nl}$ is the cyclic
frequency of a mode of order $n$ and degree $l$, and $\alpha$ is in
general a slowly varying function of frequency.  
The departures from the simple relation in Eq.~(\ref{eq:largesep}) and
Eq.~(\ref{eq:lplustwo})  provide diagnostic information about the
stellar interior.

In the simplest approximation, the modes are uniformly spaced in radial
order. The so-called ``large separation,'' $\Delta$, is approximately
constant,
\begin{equation}
\Delta(n,l) ~\simeq~ \nu_{n,l}-\nu_{n-1,l}.
\label{eq:larg}
\end{equation} 
The large separation $\Delta$ can be related to the sound speed by the
formula 
\begin{equation}
\Delta = \left(2 \int_0^R {d r \over c(r)} \right)^{-1} \; .
\label{eq:sepeq}
\end{equation}
In Eq.~(\ref{eq:sepeq}), the integration variable is the distance $r$
from the center of the star; the range of integration is from the
solar center to the surface radius $R$. The sound speed is denoted by
$c$.  According to this equation, $\Delta$ is the inverse of the
sound travel time across a stellar diameter. Thus as the the sound
speed decreases with increasing solar age, the sound travel time
increases and the large separations decrease.  It also follows from
Eq.~(\ref{eq:largesep}) that the frequencies of modes whose degrees
have the same parity approximately coincide,
\begin{equation}
\nu_{nl} ~ \simeq ~ \nu_{n-1\,l+2} \; .
\label{eq:lplustwo}
\end{equation}

The departure from Eq.~(\ref{eq:lplustwo}), the ``small frequency
separation,'' is defined by the relation
\begin{equation}
d_{nl}~ =~ \nu_{nl} - \nu_{n-1\,l+2} ,
\label{eq:small}
\end{equation}
 and is sensitive to conditions in the stellar core. The small
separation $d_{nl}$ depends on the gradient of the solar sound speed,
particularly near the stellar core. The average over $n$ of $d_{nl}$
is proportional to a quantity $D$ given by
\begin{equation}
D\simeq -{1\over 4 \pi^2 x_0}\int_0^R {dc\over dr} {dr\over r},
\label{eq:smallcont}
\end{equation}
 where $x_0$ is a suitable reference value of $n+ l/2$ (Scherrer et
al. 1983;\nocite{scherrer83} Christensen-Dalsgaard
1988).\nocite{jcd88} The average small frequency separations provide a
measure of the age of the star (see, e.g., Ulrich
1986;\nocite{ulrich86} Christensen-Dalsgaard 1988;\nocite{jcd88}
Guenther 1991).\nocite{guen91} The solar sound speed decreases with
age. The main decrease occurs near the core, in large part due to an
increase with age of the mean molecular weight as a result of hydrogen
being converted to helium. The sound-speed develops a local minimum at
the center. Thus $dc/dr$ is positive near the center of the star (in
contrast to the negative gradient elsewhere), which reduces the value
of the integral in Eq.~(\ref{eq:smallcont}). The small frequency
separations are more sensitive to age than the large frequency
separations, although the large separations also decrease with
increasing age.

Table~\ref{tab:separations} lists for our standard solar model the
calculated unweighted averages (denoted by angular brackets) of the
large and small frequencies separations as a function of solar age.
The large separations were averaged over modes with $l=0$, $1$, $2$
and $3$ and $n$ between $10$ and $22$. This range of $n$ was chosen
because it is the observed range for most data sets.  The set of small
separations given in the third column of
Table~\ref{tab:separations} ($d_{n0}$) was calculated for $l=0$ and
$l=2$ modes; the small separations in the fourth column of
Table~\ref{tab:separations} ($d_{n1}$) were calculated for $l=1$ and
$l=3$ modes. The small separations were averaged over the same range
of $n$ as the large separations. 

\begin{table}[!htb]
\begin{center}
\singlespace
\caption[]{The average large and small frequency separations of low
degree modes of a solar model as a function of solar age. Also listed
are the observed values for the the Sun.  All splittings are given in
units of $\mu$Hz.
\label{tab:separations}}
\begin{tabular}[h]{lccc}
\hline \\[-5pt]
Age (Gyr) & $\langle \Delta(n,l) \rangle$ & 
$\langle d_{n0}\rangle$& $\langle d_{n1}\rangle$ \\
\hline \\[-5pt]
     0.000     &  170.57  &     19.65     &         31.34\\
     0.547  & 164.03   &    17.84      &        28.76\\
     1.083  & 160.22   &    16.47      &        27.55\\
     1.888  & 154.87   &    15.25      &        25.08\\
     2.246 & 152.62   &    14.70      &        24.29\\
     2.871 & 147.98   &    13.42      &        22.29\\
     3.408 & 144.52   &    12.60      &        21.10\\
     4.034 & 140.16   &    11.67      &        19.68\\
     4.570   & 136.10   &    10.57      &        17.97\\
     5.017  & 132.87   &     9.92      &        16.95\\
     5.464  & 129.57   &     8.98      &        15.53\\
     5.911  & 126.19   &     8.37      &        14.50\\
     6.358  & 122.82   &     7.45      &        13.09\\
     6.805  & 119.27   &     6.90      &        12.07\\
     7.252  & 115.75   &     6.02      &        10.65\\
\hline \\[-5pt]
\multispan4{Observations\hfill}\\
BiSON+LOWL & 135.33  &   10.51   &        17.81 \\
GOLF       & 135.12  &   10.46   &        17.75 \\
GOLF2      & 134.84  &   10.22   &        17.18 \\
GONG       & 134.86  &   10.04   &        18.24 \\
MDI        & 134.95  &   10.14   &        17.64\\
\hline \\[-5pt]
\end{tabular}
\end{center}
\end{table}

As expected, both large and small separations decrease with time, with
small separations decreasing much more rapidly than large
separations. The decrease is slightly larger for the $l=0,2$
separations than for the $l=1,3$ separations because the $l=0$ modes
are more sensitive to core conditions than the $l=1$ modes.  

Table~\ref{tab:separations} also shows the
results obtained from different helioseismology
observations. References for the observations are given in
\S~\ref{sec:soundspeeds}, which is where we discuss sound speeds.
We do not include in Table~\ref{tab:separations} results for the large
and small separations for LOWL1, because for this data set there are
not enough low degree modes to form a robust average over the range of
selected $n$.

The calculated values for a solar age of $4.57$ Gyr should be compared
with the observed values listed in Table~\ref{tab:separations}.  The
calculated large separation is slightly different from the observed
values, which may reflect an uncertainty in the detailed physics of
the surface layers. The uncertainty in the physics of the surface
layers introduces errors in the calculated model frequencies. These
frequency-dependent errors are not completely removed in the process of
calculating the large separations from the standard solar model.  The
surface uncertainties are removed more successfully in the calculation
of the small separations.

\section{Variant and deviant solar models}
\label{sec:variantdeviant}

In this section, we describe eleven solar models, seven of which are
slight variants on the theme of the standard solar model (see
\S~\ref{subsec:variants}) and four of which are deficient in one or
more significant aspects of the physics used in their construction
(see \S~\ref{subsec:deviants}).  Nine of these solar models
have been described in detail in BPB2000, where their
helioseismological properties were investigated extensively.  Hence
the descriptions here will be brief.

Since the first report (Davis, Harmer, \& Hoffman
1968)\nocite{davis68} that the solar neutrino event rate in the
chlorine experiment was less than the predicted rate (Bahcall,
Bahcall, \& Shaviv 1968),\nocite{bb68} obtained by using the then
standard solar model and the then standard electroweak theory, there
have been many studies of deviant solar models that were designed to
`explain' the solar neutrino problem.  The first non-standard solar
model was proposed by Ezer \& Cameron (1968),\nocite{ezer68} who
suggested that the flux of $^8$B neutrinos could be reduced if the
central hydrogen abundance could be maintained near its initial value
by continuous mixing in the solar core. (For a discussion of, and
references to, the extensive early work on this problem, see Chapter 5
of Bahcall 1989.)  With the advent of precise measurements of solar
$p$-modes that extend deep into the solar interior, many non-standard
models have been explicitly shown to be inconsistent with the inferred
solar sound speeds (Bahcall, Pinsonneault, Basu, \&
Christensen-Dalsgaard 1997)\nocite{bp97} or the $p$-mode frequencies
(Guenther \& Demarque 1997).\nocite{guen97}

We explore here the range of solar parameters predicted by various
non-standard models, even those that are strongly inconsistent with
helioseismological data. Our purpose is to set extreme limits on
predicted solar parameters, such as the luminosity evolution or
neutrino emission, rather than the traditional goal of avoiding new
neutrino physics. Other authors have used non-standard solar models in
connection with helioseismology for a variety of different
applications, including, for example, constraining the cross section
for the $p-p$ interaction (Antia \& Chitre 1999),\nocite{antia99}
limiting the amount of mass loss from the sun (Guzik \& Cox
1995),\nocite{guzik95} and constraining the amount of anomalous energy
transport by WIMPS (Christensen-Dalsgaard 1992).\nocite{jcd92}
Turcotte \& Christensen-Dalsgaard (1998a)\cite{turcotte98a} have considered the
impact of changes in heavy element abundances on the properties of
some models; the pattern of effects they found are consistent with
those obtained here.

Table~\ref{tab:interior} and Table~\ref{tab:cz} summarize some
of the important physical characteristics of the complete set of the
nine solar models whose helioseismological properties we studied in
BPB2000 plus two models with slightly lower values of $Z/X$ that are
studied here for the first time.  We present in
Table~\ref{tab:interior} for each of the models the central
temperature, density, and pressure; the helium and heavy element mass
fractions in the initial model; the helium and heavy element mass
fractions in the solar center. We give in Table~\ref{tab:cz} the
helium and heavy element abundances at the solar surface; the mixing
length parameter, $\alpha$; and the radius and temperature at the base
of the convective zone, as well as the mass included in the convective
zone.  All quantities are shown for the model at the present epoch.

We will now define the eleven Variant and Deviant solar models.

\begin{table}[!ht]
\centering
\singlespace
\caption{\baselineskip=12pt
Some interior characteristics of the solar
models. The quantities
$T_c$ (in units of $10^7$ K), $\rho_c$ $(10^2~{\rm gm~cm^{-3}})$,
and $P_c$ $(10^{17}~{\rm erg~cm^{-3}})$ are the current epoch
central temperature, density, and
pressure; Y and Z are the helium and heavy element mass fractions, where
the subscript `init' denotes the zero-age main sequence model, and the
subscript `c' denotes the center of the solar model.
\label{tab:interior}}
\begin{tabular}{lccccccc}
\tableline\tableline
Model&$T_c$&$\rho_c$&$P_c$
&$Y_{\rm init}$&$Z_{\rm init}$&$Y_c$&$Z_c$\\
\tableline
Standard&15.696&152.7&2.342&0.2735&0.0188&0.6405&0.0198\\
NACRE&15.665&151.9&2.325&0.2739&0.0188&0.6341&0.0197\\
AS00&15.619&152.2&2.340&0.2679&0.0187&0.6341&0.0197\\
GN93& 15.729 & 152.9 & 2.342 & 0.2748 & 0.02004 & 0.6425 & 0.02110\\
Pre-M.S.& 15.725 & 152.7 & 2.339 & 0.2752 & 0.02003 & 0.6420 & 0.02109\\
Rotation& 15.652 & 148.1 & 2.313 & 0.2723 & 0.01934 & 0.6199 & 0.02032\\
${\rm Radius_{78}}$& 15.729 & 152.9 & 2.342 & 0.2748 & 0.02004 & 0.6425 &
0.02110\\
${\rm Radius_{508}}$& 15.728 & 152.9 & 2.341 & 0.2748 & 0.02004 & 0.6425 &
0.02110\\
No Diffusion& 15.448 & 148.6 & 2.304 & 0.2656 & 0.01757 & 0.6172 & 0.01757\\
Old physics& 15.787 & 154.8 & 2.378 & 0.2779 & 0.01996 & 0.6439 & 0.02102\\
$S_{34} = 0$& 15.621 & 153.5 & 2.417 & 0.2722 & 0.02012 & 0.6097 & 0.02116\\
Mixed& 15.189 & 90.68 & 1.728 & 0.2898 & 0.02012 & 0.3687 & 0.02047\\
\noalign{\smallskip}
\tableline
\end{tabular}
\end{table}

\begin{table}[!ht]
\centering
\singlespace
\caption{\baselineskip=12pt
Some characteristics of the convective zones of solar
models at the current epoch. Here $Y_s$ and $Z_s$ are the surface
helium and heavy element abundances, $\alpha$ is the mixing length
parameter, $R({\rm
CZ})$ and $T({\rm CZ})$ are the radius and temperature
at the base of the convective zone, and $M({\rm CZ})$ is
the mass included within
the convective zone.
\label{tab:cz}}
\begin{tabular}{lcccccc}
\tableline\tableline
Model&$Y_s$&$Z_s$&$\alpha$&$R{(\rm CZ)}$&$M{(\rm CZ)}$
&$T({\rm CZ})$\\
&&&&$(R_\odot)$&$(M_\odot)$&$(10^6~{\rm K})$\\
\tableline
Standard&0.2437&0.01694&2.04&0.7140&0.02415&2.18\\
NACRE&0.2443&0.01696&2.04&0.7133&0.02451&2.19\\
AS00&0.2386&0.01684&2.05&0.7141&0.02394&2.18\\
GN93&0.2450 & 0.01805 & 2.06 & 0.7124 & 0.02457 & 2.20\\
Pre-M.S.&0.2455 & 0.01805 & 2.05 & 0.7127 & 0.02443 & 2.20\\
Rotation&0.2483 & 0.01797 & 2.03 & 0.7144 & 0.02388 & 2.15\\
${\rm Radius_{78}}$& 0.2450 & 0.01806 & 2.06 & 0.7123 & 0.02461 & 2.20\\
${\rm Radius_{508}}$& 0.2450 & 0.01806 & 2.06 & 0.7122 & 0.02467 & 2.20\\
No Diffusion&0.2655 & 0.01757 & 1.90 & 0.7261 & 0.02037 & 2.09\\
Old physics&0.2476 & 0.01796 & 2.04 & 0.7115 & 0.02455 & 2.21\\
$S_{34} = 0$&0.2422 & 0.01811 & 2.03 & 0.7151 & 0.02309 & 2.17\\
Mixed& 0.2535 &
0.01782 & 1.85 & 0.7315 & 0.01757 & 2.02\\
\noalign{\smallskip}
\tableline
\end{tabular}
\end{table}

\subsection{Variant solar models}
\label{subsec:variants}

The NACRE model was constructed using the same input physics as our
Standard model except that we use for the NACRE model the charged
particle fusion cross sections recommended in the NACRE compilation
(Angulo et al. 1999).\nocite{angulo99} We specify the fusion cross
sections used for this model more fully in \S~\ref{subsubsec:nacre}.
The model GN93 was considered our Standard model in BPB2000 and
differs only in the adopted value of $Z/X = 0.0245$ (Grevesse \& Noels
1993)\nocite{gre93} from the current Standard model (see
\S~\ref{sec:standard}) which has $Z/X = 0.0230$ (Grevesse \& Sauval
1998).\nocite{gre98} The model AS00 is the same as the two models
described above except that it has lower heavy element abundance $Z/X
= 0.0226$ (Asplund 2000).\nocite{asplund00} As a consequence of a more
detailed calculation of the solar atmosphere, Asplund
(2000)\nocite{asplund00} suggests that all meteoritic abundances
should be adjusted downward by 0.04 dex.  All of the models described
below, in \S~\ref{subsec:deviants} as well as in this subsection, use
the Grevesse and Noels (1993)\nocite{gre93} composition mix with $Z/X
= 0.0245$.

Model Pre-M.S. is evolved from the
pre-main sequence stage, but otherwise is the same as our Standard model.
The model Rotation incorporates mixing induced by rotation and is a
reasonable upper bound to the degree of rotational mixing which is
consistent with the observed depletion of lithium in the sun
\footnote{The Rotation model discussed here differs somewhat from the
rotation model analyzed in BPB2000 in that the metals heavier than CNO
were inadvertently not mixed in the previous version of this model.
The rotation profile computed from this model does not match precisely
the best current estimates of the rotation profile in the inner
regions of the sun.  The case considered here corresponds to the
maximum amount of mixing. Richard et al. (1996)\nocite{ri96} consider
solid body rotation, which corresponds to what is likely to be the
minimum amount of mixing.  Both of these models yield similar results
for the effect of rotation on diffusion, which is the principal way
that rotation affects solar neutrino fluxes.}(Pinsonneault et
al. 1999).\nocite{marc99} The prescriptions for calculating this model
are described in \S~5 of Pinsonneault (1997)\nocite{marc97} and in
BPB2000.  There has been considerable discussion recently regarding
the precise value of the solar radius (cf. Antia 1998;\nocite{a98}
Schou et al. 1997;\nocite{sch97} Brown \& Christensen-Dalsgaard
1998)\nocite{br98} and some discussion of the effects of the
uncertainty in radius on the quantities inferred from the
helioseismological inversions (cf. Basu 1998).\nocite{GONGMDI} We have
therefore considered two models which were constructed with the same
input physics as BP2000, but which have model radii which differ from
the radius assumed in constructing the Standard model.  ${\rm
Radius_{78}}$ has a radius of 695.78 Mm, which is the radius that has
been determined from the frequencies of f-modes (cf. Antia
1998)\nocite{a98} and ${\rm Radius_{508}}$ has a radius of 695.508 Mm,
which is the solar radius as determined by Brown \&
Christensen-Dalsgaard (1998),\nocite{br98} who used the measured
duration of solar meridian transits during the 6 years 1981--1987 and
combined these measurements with models of the solar limb-darkening
function to estimate the value of the solar radius.

All of these variant models are approximately as consistent with the
helioseismological evidence as the Standard model (see BPB2000).  For
example, the rms sound speed differences between the variant models
and BP2000 are: $0.03$\% (Pre-M.S.), $0.08$\% (Rotation), $0.15$\%
(Radius$_{78}$), and $0.03$\% (Radius$_{508}$).  The average
difference (rms) between the four variant models and the Standard
model is $0.07$\% . We shall see in \S~\ref{sec:neutrinos} that the
differences predicted by these models for the important neutrino
fluxes are all less than $5$\%.

\subsection{Deviant solar models}
\label{subsec:deviants}

The model `Old physics' was constructed using the old Yale equation of
state (cf. Guenther et al. 1992),\nocite{guen92} supplemented with the
Debye-H\"uckel correction (cf. Bahcall, Bahcall, \& Shaviv
1968)\nocite{bb68} and older OPAL radiative opacities (Iglesias,
Rogers, \& Wilson 1992;\nocite{ig92} Kurucz 1991).\nocite{kur91} The
model includes helium and heavy element diffusion and nuclear reaction
cross section data in the same way as our Standard model.  The S$_{34}
= 0$ model was calculated assuming that the rate of the
$^3$He$(\alpha,\gamma)^7$Be reaction is zero, which implies that no
$^7$Be or ${}^8$B neutrinos are produced.  In the Standard solar
model, about $12$\% of the terminations of the $p-p$ chain involve the
$^3$He$(\alpha,\gamma)^7$Be reaction, whose rate is proportional to
S$_{34}$.  The No Diffusion model does not include helium or
heavy-element diffusion and therefore represents the state-of-the art
in solar modeling prior to 1992 (cf.  Bahcall \& Ulrich
1988;\nocite{bu88} Bahcall \& Pinsonneault 1992;\nocite{bp92} Proffitt
1994).\nocite{proffitt94} The model Mixed has an artificially mixed
core, with the inner $50$\% by mass ($25$\% by radius) required to be
chemically homogeneous at all times.  This model was constructed to be
similar to the prescription of Cumming \& Haxton (1996),\nocite{chx96}
who changed by hand the $^3$He abundance as a function of radius in
the final BP95 (Bahcall \& Pinsonneault 1995)\nocite{bp95} solar model
in order to minimize the discrepancy between measurements of the total
event rates in neutrino experiments and the calculated event
rates. Cumming and Haxton did not calculate the time evolution of
their model.

We showed in BPB2000 that the Mixed, No Diffusion, and $S_{34} =0$
models are strongly disfavored by helioseismological data. We use
these deviant (or deficient) models here to test the robustness of the
discrepancies between solar model predictions and solar neutrino
measurements. (For references and discussions to earlier work on these
deviant models, see Bahcall 1989,\nocite{bahcall89}
Christensen-Dalsgaard 1995,\nocite{jcd95} Bahcall, Pinsonneault, Basu,
\& Christensen-Dalsgaard 1997,\nocite{bp97}, Guenther \& Demarque
1997\nocite{guen97}, and Cox, Kidman, and Newman 1985\nocite{cox85}). We
have already seen in Figure~\ref{fig:lumnormalized} that the
luminosity evolution predicted by solar models is essentially the same
for the deviant models and for the Standard solar model.

\section{Neutrino Physics}
\label{sec:neutrinos}
This section presents neutrino fluxes and predicted event rates in
different solar neutrino experiments; the fluxes and event rates are
calculated using the eleven solar models described in
\S~\ref{sec:variantdeviant}. We also give  other aspects of the solar model that
influence the interpretation of solar neutrino experiments.

As described in \S~\ref{subsec:standardcharacteristics}, we present
at http://www.sns.ias.edu/$\sim$jnb a detailed numerical table which
gives the fraction of each of the eight important neutrino fluxes that
is produced in each spherical shell. These neutrino production
fractions are important for calculating the effect of MSW (matter)
oscillations in the sun, but will not be discussed further here.

Section~\ref{subsec:fluxestoday} presents the fluxes of electron type
neutrinos that are produced in the sun according to the Standard solar
model and to the eight variant solar models considered in this
paper. We compare the results of these predictions with measurements
of the total rates in the chlorine solar neutrino experiment, the SAGE
and GALLEX + GNO gallium experiments, and the Kamiokande and
Super-Kamiokande water Cherenkov detectors. We also describe in this
section how we estimate the errors (asymmetric in some cases) on the
predicted fluxes and event rates.

In \S~\ref{subsec:electrondensity}, we present detailed results for
the electron number density versus solar radius. The MSW conversion of
electron type neutrinos to other neutrino types depends upon the
radial distribution of the electron number density. In previous
presentations of the Standard model, we have not given the electron
number density in the outer parts of the sun with sufficient precision
to permit calculations of a subset of the currently allowed matter
transitions.
Section~\ref{subsec:fluxesvsage} presents the calculated
time-dependences of the most important solar neutrino fluxes, the
$pp$, $^7$Be, $^8$B, and $^{13}$N neutrino fluxes.

\subsection{Neutrino fluxes and experimental event rates at the current epoch}
\label{subsec:fluxestoday}
We present  in \S~\ref{subsubsec:standardneutrino} the neutrino fluxes
and experimental event rates predicted by the Standard model and
contrast these results with the observed rates. In
\S~\ref{subsubsec:uncertainties}, we describe the procedures and
the ingredients used to calculate the uncertainties in the neutrino
fluxes and the event rates. We compare  in
\S~\ref{subsubsec:variantdeviantneutrino} the calculated neutrino
fluxes and experimental event rates for the eight variant and deviant
solar models with the results for the Standard model and with the
measured solar neutrino event rates.

\subsubsection{Standard model}
\label{subsubsec:standardneutrino}

\begin{table}[!htb]
\centering
\singlespace
\caption[]{\baselineskip=12pt Standard Model Predictions (BP2000):
solar neutrino fluxes and neutrino capture rates, with $1\sigma$
uncertainties from all sources (combined quadratically).  The
tablulated fluxes correspond to a present-day solar luminosity of $
3.842 \times 10^{33} {\rm ~erg s^{-1}}$.  The observed capture rates
are: $2.56 \pm 0.23$ SNU [chlorine (Lande 2000)]\nocite{lande00} and
$74.7 \pm 5.0 $ SNU [combined SAGE and GALLEX plus GNO (Hampel et
al. 1999;\nocite{hampel99} Abdurashitov et al. 1999;\nocite{abdur99}
Bellotti 2000)].\nocite{belotti2000} The $^8$B flux measured by the
Super-Kamiokande experiment is $2.40 \pm 0.03 ({\rm stat})
~^{+0.08}_{-0.07} ({\rm syst.})\,\,{\rm cm^{-2}s^{-1}}$ (Suzuki 2000).
\nocite{suzuki2000} \protect\label{tab:bestestimate}}
\begin{tabular}{llccc}
\tableline\tableline
Source&\multicolumn{1}{c}{Flux}&Cl&Ga&Li\\
&\multicolumn{1}{c}{$\left(10^{10}\ {\rm cm^{-2}s^{-1}}\right)$}&(SNU)&(SNU)&(SNU)\\
\tableline
pp&$5.95 \left(1.00^{+0.01}_{-0.01}\right)$&0.0&69.7&0.0 \\
pep&$1.40 \times 10^{-2}\left(1.00^{+0.015}_{-0.015}\right)$&0.22&2.8&9.2 \\
hep&$9.3 \times 10^{-7}$&0.04&0.1&0.1 \\
${\rm ^7Be}$&$4.77 \times
10^{-1}\left(1.00^{+0.10}_{-0.10}\right)$&1.15&34.2&9.1 \\
${\rm ^8B}$&$5.05 \times
10^{-4}\left(1.00^{+0.20}_{-0.16}\right)$&5.76&12.1&19.7  \\
${\rm ^{13}N}$&$5.48 \times
10^{-2}\left(1.00^{+0.21}_{-0.17}\right)$&0.09&3.4&2.3 \\
${\rm ^{15}O}$&$4.80 \times
10^{-2}\left(1.00^{+0.25}_{-0.19}\right)$&0.33&5.5&11.8 \\
${\rm ^{17}F}$&$5.63 \times
10^{-4}\left(1.00^{+0.25}_{-0.25}\right)$&0.0&0.1&0.1 \\
\noalign{\medskip}
Total&&$7.6^{+1.3}_{-1.1}$&$128^{+9}_{-7}$&$52.3^{+6.5}_{-6.0}$ \\
\tableline
\end{tabular}
\tablecomments{The cross sections for neutrino absorption on chlorine are from
Bahcall et al. (1996);\nocite{bl96} the cross sections for 
gallium are from Bahcall
(1997);\nocite{bahcall97} the cross sections for ${\rm ^7Li}$ are from Bahcall (1989)\nocite{bahcall89}
and Bahcall (1994).\nocite{bahcall94}}
\end{table}

Table~\ref{tab:bestestimate} gives the neutrino fluxes and their
uncertainties for our Standard solar model. In order to obtain the
most precise values that we can for the predicted fluxes, we have
recomputed the Standard model discussed elsewhere in this paper. We
use in this subsection the most recently determined absolute value for
the solar luminosity, $ 3.842 \times 10^{33} {\rm ~erg s^{-1}}$
(Fr\"ohlich \& Lean 1998\nocite{frohlich98}, Crommelynck, Fichot,
Domingo, \& Lee 1996\nocite{crommel96}), which is $0.2$\% smaller 
than the value used in the model calculations discussed elsewhere in
this paper. The largest changes due to the $0.2$\% decrease in the solar
luminosity are a $2$\% decrease in the $^8$B neutrino flux and a 
$1$\% decrease in the $^7$Be neutrino flux.  All other quantities
calculated in this paper are changed by negligible amounts.

The adopted uncertainties in different input parameters are given in
Table~2 of Bahcall, Basu, \& Pinsonneault (1998),\nocite{ba98} which
we refer to hereafter as BP98. We also present in
Table~\ref{tab:bestestimate} the calculated event rates in the
chlorine, gallium, and lithium experiments.  The rate in the
Super-Kamiokande experiment is usually quoted as a fraction of the
best-estimate theoretical flux of $^8$B neutrinos, assuming an
undistorted (standard) energy spectrum.

\begin{table}[!htb]
\centering
\singlespace
\caption[]{\baselineskip=12pt Solar Neutrino Rates: Theory versus
Experiment.  The units are SNU ($10^{-36}$ interactions per atom per
sec) for the radiochemical experiments: Chlorine, GALLEX + GNO, and
SAGE.  The unit for the ${\rm ^8B}$ and hep fluxes are, respectively,
$10^6~{\rm cm^{-2}~s^{-1}}$ and $10^3~{\rm cm^{-2}~s^{-1}}$. The
errors quoted for Measured/BP2000 are the quadratically combined
uncertainties for both BP2000 and the Measured rates.  For simplicity
in presentation, asymmetric errors were averaged.  References to the
experimental results are given in the text and in Lande
(2000),\nocite{lande00} Bellotti (2000),\nocite{belotti2000} Gavrin
(2000),\nocite{gavrin2000} Fukuda et al. (1996),\nocite{fukuda96} and
Suzuki (2000)\nocite{suzuki2000} for the chlorine, GALLEX + GNO, SAGE,
Kamiokande, and Super-Kamiokande results.
\label{tab:numeasurements}}
\begin{tabular}{lccc}
\tableline\tableline
\multicolumn{1}{c}{Experiment}&BP2000&Measured&Measured/BP2000\\
\tableline
Chlorine&$7.6^{+1.3}_{-1.1}$&$2.56 \pm 0.23$&$0.34 \pm 0.06$\\
GALLEX + GNO&$128^{+9}_{-7}$&$74.1^{+6.7}_{-7.8}$&$0.58 \pm 0.07$\\
SAGE&$128^{+9}_{-7}$&$75.4^{+7.8}_{-7.4}$&$0.59 \pm 0.07$\\
${\rm ^8B}$-Kamiokande&$5.05\left[1.00+
^{+0.20}_{-0.16}\right]$&$2.80\left[1.00 \pm 0.14\right]$&$0.55
\pm 0.13$\\
${\rm ^8B}$-Super-Kamiokande&$5.05 \left[1.00 +
^{+0.20}_{-0.16}\right]$&$2.40 \left[1.00 +
^{+0.04}_{-0.03}\right]$&$0.48 \pm 0.09$\\
hep-Super-Kamiokande&9.3&$11.3(1 \pm 0.8)$&$\sim 1$\\
\tableline
\end{tabular}
\end{table}

Table~\ref{tab:numeasurements} compares the predictions of the
combined standard model, i. e., the standard solar model (BP2000) and
the standard electroweak theory (no neutrino oscillations), with the
results of the chlorine, GALLEX + GNO, SAGE, Kamiokande, and
Super-Kamiokande solar neutrino experiments.  The observed rate in the
chlorine experiment is $2.56 \pm 0.23$ SNU (Lande
2000;\nocite{lande00} Davis 1994;\nocite{davis94} Cleveland et
al. 1998),\nocite{cleveland98} which is to be compared to the
calculated value of $7.6^{+1.3}_{-1.1}$ SNU. This discrepancy between
calculated and observed neutrino capture rates has been approximately
the same for more than three decades (cf. Bahcall, Bahcall, \& Shaviv
1968;\nocite{bb68} Davis, Harmer, \& Hoffman 1968;\nocite{davis68}
Bahcall 1989).\nocite{bahcall89}

The average of the SAGE (Abdurashitov et al. 1999;\nocite{abdur99} 
Gavrin 2000\nocite{gavrin2000}) and
the GALLEX (Hampel et al. 1999)\nocite{hampel99} plus GNO 
(Bellotti 2000)\nocite{belotti2000} results is
$74.7 \pm 5.0$ SNU, which is more than $6\sigma$ away from the
calculated standard rate of $128^{+9}_{-7}$ SNU.

After $1117$ days of data acquisition, the flux of $^8$B neutrinos
measured by Super-Kamiokande is $[2.40 \pm 0.03 ({\rm stat})
~^{+0.08}_{-0.07} ({\rm syst.})] 10^{-6}\,\,{\rm cm^{-2}s^{-1}}$
(Suzuki 2000;\nocite{suzuki2000} Fukuda et al. 1998),\nocite{fukuda98}
which corresponds to $0.475$ of the BP2000 predicted flux.

Comparing the second and third columns of
Table~\ref{tab:numeasurements}, we see that the predictions of the
combined standard model differ by many standard deviations from the
results of the solar neutrino experiments.

The flux of $hep$ neutrinos was calculated for BP2000 using the most
recent theoretical evaluation by Marcucci et
al. (2000a,b)\nocite{marcucci00a,marcucci00b} of the cross section
factor $S_0({\rm hep})$, which is $4.4$ times larger than the previous
best estimate. The most recent preliminary report (after $1117$ days
of data) of the Super-Kamiokande collaboration (Suzuki
2000)\nocite{suzuki2000} is that the $hep$ flux observed in their
$\nu_e-e$ scattering experiment is $5.4 \pm 4.5$ times the
best-estimate from BP98. Since the BP2000 estimate of the $hep$ flux
is a factor of $4.4$ times larger than the flux quoted in BP98, the
best-estimate theoretical $hep$ flux now agrees with the best-estimate
experimental $hep$ flux measurement, although we do not attach much
significance to this agreement since we cannot quote an uncertainty on
the theoretical estimate (see discussion of the $hep$ reaction in
\S~\ref{subsubsec:uncertainties}).

The event rates predicted by BP2000 for the chlorine, gallium, and
Super-Kamiokande solar neutrino experiments are within two percent of
the rates predicted for the BP98 standard solar models.  As far as
these experiments are concerned, the effects of using the improved
heavy element composition essentially cancels the effect of correcting
the error in the opacity interpolation (see
\S~\ref{subsec:definitionstandard}). The difference in the $^7$Be flux
predicted by BP98 and BP2000 is only $0.6$\%.  The $^7$Be flux will be
measured by the BOREXINO solar neutrino experiment.

\subsubsection{Calculated uncertainties}
\label{subsubsec:uncertainties}

We have calculated the uncertainties in the neutrino fluxes and in the
experimental event rates by including the published errors in all
experimental quantities and by taking account of the correlated
influence of different input parameters using the results of detailed
solar model calculations.  The procedure for calculating the
uncertainties has been developed over the past three decades and is
described in detail in Chapter~7 of Bahcall (1989)\nocite{bahcall89}
(see also Bahcall \& Pinsonneault 1992, 1995,\nocite{bp92,bp95} and
Bahcall, Basu, \& Pinsonneault 1998).\nocite{ba98}

In order that the reader can see the specific implementation of the
uncertainty calculations, we are making available the exportable
fortran code that evaluates the rates in different neutrino
experiments and also calculates the uncertainties in the individual
neutrino fluxes and experimental rates. The code, exportrates.f, is
available at http://www.sns.ias.edu/$\sim$jnb/SNdata/sndata.html .

The uncertainties in the nuclear fusion cross sections (except for
$hep$, see below) were taken from Adelberger et
al. (1998),\nocite{adel98} 
the neutrino
cross sections and their  uncertainties are from Bahcall (1994, 1997)
\nocite{bahcall94,bl96,bahcall97} 
and Bahcall et al. (1996), the luminosity and age uncertainties were
adopted from Bahcall \& Pinsonneault (1995),\nocite{bp95} 
the $1\sigma$ fractional
uncertainty in the diffusion rate was taken to be $15$\% (Thoul,
Bahcall, \& Loeb 1994)\nocite{thoul4} and the opacity uncertainty was determined by
comparing the results of fluxes computed using the older Los Alamos
opacities with fluxes computed using the modern Livermore opacities
(Bahcall \& Pinsonneault 1992).\nocite{bp92}

We follow the discussion in Bahcall \& Pinsonneault (1995)\nocite{bp95} and adopt a
$1 \sigma$ uncertainty in the heavy element abundance of
\begin{equation}
\sigma (Z/X) = \pm 0.061 \times (Z/X).
\label{eqn:zoverx}
\end{equation}
This uncertainty spans the range of values recommended by Grevesse
(1984),\nocite{gre84} Grevesse \& Noels (1993),\nocite{gre93} and
Grevesse \& Sauval (1998)\nocite{gre98} over the fourteen year period
covered by the cited Grevesse et al. review articles.  The uncertainty
adopted here is about twice as large as the uncertainty recommended by
Basu \& Antia (1997)\nocite{basu97} based on their helioseismological
analysis.  In support of the larger uncertainty used in this paper, we
note that the difference between the Grevesse \& Noels
(1993)\nocite{gre93} values of $Z/X = 0.0230$ is $1 \sigma$ according
to Eq.~(\ref{eqn:zoverx}).

We include for the first time in this series of papers the uncertainty
in the small ${\rm ^{17}F}$ neutrino flux due to the uncertainty in the
measured $S_0$-factor for the reaction ${\rm ^{16}O}(p,\gamma){\rm
^{17}F}$.  We use the $1\sigma$ uncertainty 18.1\% estimated by
Adelberger et al. (1998).\nocite{adel98}  
It was an oversight not to include this
uncertainty in our previous calculations.

The only flux for which we do not quote an estimated uncertainty is
the $hep$ flux (see Table~\ref{tab:bestestimate}). The difficulty of
calculating from first principles the nuclear cross section factor $S_0({\rm hep})$ is what
has caused us not to quote in this series of papers an uncertainty in
the $hep$ flux (see discussion in Bahcall 1989\nocite{bahcall89} 
and in Bahcall \& Krastev
1998).\nocite{bak98}  The $hep$ reaction is uniquely 
difficult to calculate among
the light element fusion reactions since the one-body and two-body
contributions to the reaction rate are comparable in magnitude but
opposite in sign, so that the net result is sensitive to a delicate
cancellation. Also, two-body axial currents from excitations of
$\Delta$ isobars are model dependent. In addition, the calculated rate
is sensitive to small components in the wave function, particularly
$D$-state admixtures generated by tensor interactions. These
complications have been discussed most recently and most thoroughly by
Marcucci et al. (2000b).\nocite{marcucci00b}

The calculated errors are asymmetric in many cases. These asymmetries
in the uncertainties in the neutrino fluxes and experimental event
rates result from asymmetries in the uncertainties of some of the input
parameters, for example, the important $pp$, \hbox{$^7$Be + p}, and $^{14}$N
+ p fusion reactions and the effect of excited states on neutrino
absorption cross sections.  To include the effects of asymmetric
errors, the code exportrates.f was run with different input
representative uncertainties and the different higher (lower) rates
were averaged to obtain the quoted upper (lower) limit uncertainties.

\subsubsection{NACRE charged particle fusion rates}
\label{subsubsec:nacre}

In order to estimate the systematic uncertainties associated with
different treatments of the nuclear fusion reactions, we have
constructed a solar model that is the same as the Standard Model
discussed in \S~\ref{subsubsec:standardneutrino}, except that we
have used the charged particle fusion cross sections recommended in
the NACRE compilation (Angulo et al. 1999)\nocite{angulo99} 
rather than the fusion cross
sections determined by Adelberger et al. (1998).\nocite{adel98} 
We will refer to this
solar model as the NACRE model.

The low energy cross section factors, $S_0$, that are recommended by
the NACRE collaboration and by Adelberger et al. agree within their
stated $1\sigma$ uncertainties for all of the fusion reactions that
are important for constructing a solar model.  The only important
solar nuclear reactions for which the NACRE collaboration did not
recommend interaction rates are the electron capture reactions that
produce the $^7$Be and the $pep$ neutrinos; the NACRE collaboration
also did not provide energy derivatives for the cross section factors
of the CNO reactions. Wherever the data necessary for computing solar
fusion rates was not available in the NACRE compilation, we continued
to use the Adelberger et al. (1998)\nocite{adel98} 
recommended values in computing
the NACRE model.

\begin{table}[!htb]
\centering
\singlespace
\caption[]{\baselineskip=12pt Neutrinos with NACRE reaction rates. The
solar neutrino fluxes and neutrino capture rates that were calculated
with the NACRE fusion cross sections (Angulo et
al. 1999)\nocite{angulo99} are given in
the table. The only difference between the model used in this
calculation and the Standard Model, whose fluxes are given in
Table~\ref{tab:bestestimate}, is that the Adelberger et al. (1998)\nocite{adel98}
fusion cross sections were replaced by the NACRE cross sections for
all reactions for which the NACRE collaboration quoted zero-energy
cross section factors, $S_0$.  \protect\label{tab:nacre}}
\begin{tabular}{llccc}
\tableline\tableline
Source&\multicolumn{1}{c}{Flux}&Cl&Ga&Li\\
&\multicolumn{1}{c}{$\left(10^{10}\ {\rm cm^{-2}s^{-1}}\right)$}&(SNU)&(SNU)&(SNU)\\
\tableline
pp&$5.96 \left(1.00^{+0.01}_{-0.01}\right)$&0.0&69.8&0.0 \\
pep&$1.39 \times 10^{-2}\left(1.00^{+0.015}_{-0.015}\right)$&0.22&2.8&9.1 \\
hep&$9.4 \times 10^{-7}$&0.04&0.1&0.1 \\
${\rm ^7Be}$&$4.81 \times
10^{-1}\left(1.00^{+0.10}_{-0.10}\right)$&1.15&34.5&9.1 \\
${\rm ^8B}$&$5.44 \times
10^{-4}\left(1.00^{+0.20}_{-0.16}\right)$&6.20&13.1&21.2  \\
${\rm ^{13}N}$&$4.87 \times
10^{-2}\left(1.00^{+0.21}_{-0.17}\right)$&0.08&2.9&2.1 \\
${\rm ^{15}O}$&$4.18 \times
10^{-2}\left(1.00^{+0.25}_{-0.19}\right)$&0.28&4.7&10.3 \\
${\rm ^{17}F}$&$5.30 \times
10^{-4}\left(1.00^{+0.25}_{-0.25}\right)$&0.0&0.1&0.1 \\
\noalign{\medskip}
Total&&$8.0^{+1.4}_{-1.1}$&$128^{+9}_{-7}$&$52.0^{+6.5}_{-5.9}$ \\
\tableline
\end{tabular}
\tablecomments{The cross sections for neutrino absorption on chlorine
are from Bahcall et al. (1996);\nocite{bl96} the cross sections for
gallium are from Bahcall (1997);\nocite{bahcall97} the cross sections
for ${\rm ^7Li}$ are from Bahcall (1989)\nocite{bahcall89} and Bahcall
(1994).\nocite{bahcall94}}
\end{table}

Table~\ref{tab:nacre} gives the calculated neutrino fluxes and capture
rates predicted by the NACRE solar model. In all cases, the fluxes for
the NACRE solar model agree with the fluxes calculated with the
Standard solar model to well within the $1\sigma$ uncertainties in the
Standard Model fluxes. The $^7$Be flux from the NACRE model is $1$\%
larger than for the Standard model and the $^8$B flux is $8$\%
higher. The chlorine capture rate predicted by the NACRE model is
$5$\% higher than for the Standard model; the predicted rates for the
NACRE model and the Standard model differ by less than $1$\% for the
gallium and lithium experiments.

We conclude that this estimate of the systematic uncertainties due to
the relative weights given to different determinations of nuclear
fusion cross sections suggests likely errors from this source that are
significantly smaller than our quoted $1\sigma$ errors for the
Standard model neutrino fluxes and event rates
(cf. Table~\ref{tab:bestestimate}).  Similar conclusions have been
reached by Morel, Pichon, Provost, \& Berthomieu
(1999)\nocite{morel99} in an independent investigation (see also
Castellani et al. 1997\nocite{cast97}).

\subsubsection{Variant and deviant models}
\label{subsubsec:variantdeviantneutrino}

\begin{table}[!ht]
\centering
\singlespace
\caption[]{\baselineskip=12pt
Neutrino fluxes from twelve   solar models.
The table lists the calculated neutrino fluxes and event rates
for the seven variant solar models that are discussed in
\S~\ref{subsec:variants} and for the four deficient solar models
that are discussed in \S~\ref{subsec:deviants}, and compares the
results with the Standard model fluxes. The models in rows two
through eight are all variants on the Standard model (first row).
The last four models are deficient in some important aspect of
the physics used in their construction and do not provide good
fits to the helioseismological data. \label{tab:fluxesnine}}
\begin{tabular}{lcccccccccc}
\tableline\tableline
\multicolumn{1}{c}{Model}&pp&pep&hep&${\rm ^7Be}$&${\rm ^8B}$&${\rm ^{13}N}$
&${\rm ^{15}O}$&${\rm ^{17}F}$&Cl&Ga\\
&(E10)&(E8)&(E3)&(E9)&(E6)&(E8)&(E8)&(E6)&(SNU)&(SNU)\\
\tableline
Standard&5.96&1.40&9.3&4.82&5.15&5.56&4.88&5.73&7.7&129\\
NACRE&5.97&1.39&9.4&4.85&5.54&4.93&4.24&5.39&8.1&129\\
AS00&5.99&1.41&9.4&4.62&4.70&5.25&4.56&5.33&7.1&126\\
GN 93& 5.94 & 1.39 & 9.2 & 4.88 & 5.31 & 6.18 & 5.45 & 6.50 & 8.0 &
130 \\
Pre-M.S.& 5.95 & 1.39 & 9.2 & 4.87 & 5.29 & 6.16 & 5.43 & 6.47 & 7.9 &
130 \\
Rotation& 5.98 & 1.40 & 9.2 & 4.68 & 4.91 & 5.57 & 4.87 & 5.79 & 7.4 &
127 \\
${\rm Radius_{78}}$& 5.94 & 1.39 & 9.2 & 4.88 & 5.31 & 6.18 & 5.45 & 6.50 &
8.0 & 130 \\
${\rm Radius_{508}}$& 5.94 & 1.39 & 9.2 & 4.88 & 5.31 & 6.18 & 5.45 & 6.50 &
8.0 & 130 \\
No Diffusion& 6.05 & 1.43 & 9.6 & 4.21 & 3.87 & 4.09 & 3.46 & 4.05 & 6.0 &
120 \\
Old physics& 5.95 & 1.41 & 9.2 & 4.91 & 5.15 & 5.77 & 5.03 & 5.92 & 7.8 &
130 \\
$S_{34} = 0$& 6.40 & 1.55 & 10.1 & 0.00 & 0.00 & 6.47 & 5.64 & 6.70 & 0.8 &
89 \\
Mixed& 6.13 & 1.27 & 6.2 & 3.57 & 4.13 & 3.04 & 3.05 & 3.61 & 6.1 & 115\\
\tableline
\end{tabular}
\end{table}

Table~\ref{tab:fluxesnine} compares the calculated neutrino
fluxes for the seven variant solar models described in
\S~\ref{subsec:variants} and for the four deficient solar models
described in \S~\ref{subsec:deviants} with the fluxes obtained
for the Standard solar model. The range of the $^8$B neutrino
fluxes among the seven standard-like models (rows 1--7 of
Table~\ref{tab:fluxesnine}) is only $\pm 7$\%, a factor of two
or  three smaller than the uncertainty (due to other sources) in
the calculated ${\rm ^8B}$ flux of the Standard model (see
Table~\ref{tab:bestestimate}). The spread in ${\rm ^7Be}$ flux is
$\pm 3\%$; the range in ${\rm ^{37}Cl}$ and ${\rm ^{71}Ga}$ rates
is $\pm 0.45$ SNU and $\pm 2$ SNU, respectively.

The deviant models listed in Table~\ref{tab:fluxesnine} are all
deficient in some important aspect of the physics used in their
calculation.  The No Diffusion, Old Physics, $S_{34} = 0$, and Mixed
models all give such bad agreement with helioseismology (compared to
the first seven models) that the comparison with the data cannot be
made on the same scale as for the standard-like models (see
BPB2000). The Old Physics model gives a rms difference between the
calculated and measured sound speeds that is more than twice as large
as when the Standard model is used. For the $S_{34} = 0$ model, the
rms discrepancy is seven times worse in the solar interior than for
the standard model; for the No Diffusion model, the disagreement is
about seven times worse averaged over the sun than for the standard
model. The Mixed model is the worst of all; the rms disagreement in
the solar core is about $25$ times larger than for the standard solar
model.

Even if one is willing to consider solar models that predict sound
speeds which deviate so drastically from the measured
helioseismological values, the four deviant solar models do not describe well
the solar neutrino data. For the Old Physics, No Diffusion, and Mixed
solar models, the predicted rates for the gallium experiments lie
between $115$ SNU and $130$ SNU, which is to be contrasted with the
observed rate of $75 \pm 5$ SNU, which is at least an $8\sigma$ discrepancy.  For the $S_{\rm 34} = 0$ model, the
predicted rate in the chlorine experiment is $0.79$ SNU, which is also about
eight standard deviations less than the observed value, $2.56 \pm
0.23$ SNU.

\subsection{The electron number density}
\label{subsec:electrondensity}

The probability of converting an electron type neutrino to a
muon or tau neutrino in the Sun depends upon
the profile of the electron number
density as a function of solar radius. For particular values of the
electron density, neutrino energy, and neutrino mass, neutrinos can be
resonantly converted from one type of neutrino to another. The
Mikheyev-Smirnov resonance occurs if the electron
density at a radius $r$ satisfies
\begin{equation}
\frac{n_{e,res}(r)}{N_A} ~\approx~66 \cos{2\theta_V}\left(\frac{|\Delta
m^2|}{10^{-4}~{\rm eV}}\right)\left(\frac{10 ~ {\rm MeV}}{E}\right),
\label{eq:mswresonance}
\end{equation}
where $n_e$ is the electron number density measured in ${\rm
cm^{-3}}$, $N_A$ is Avogadro's number, $\theta_V$ is the neutrino mixing
angle in vacuum, $|\Delta m^2|$ is the absolute value of the
difference in neutrino masses between two species that are mixing by
neutrino oscillations, and $E$ is the neutrino energy.

\begin{figure}[!ht]
\centerline{\psfig{figure=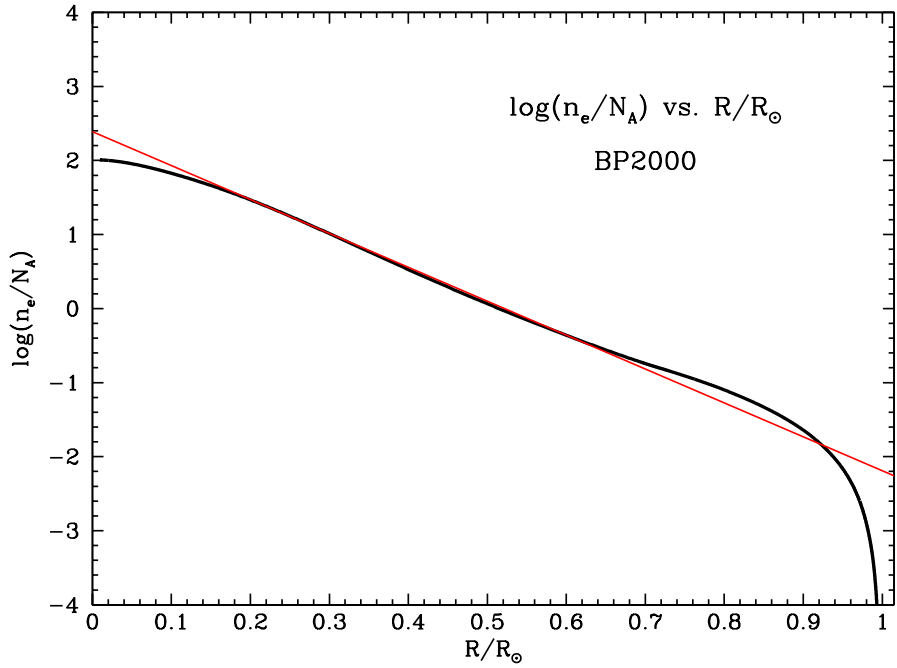,width=5in,angle=270}}
\caption{\baselineskip=12pt
The electron number density, $n_e$, versus solar radius for
the Standard solar model (BP2000). The straight-line 
fit shown in Figure~\ref{fig:ne} is
an approximation, Eq. (\ref{eq:nestraight}), given by 
Bahcall (1989).
Equation (\ref{eq:nestraight}) has been used previously in many
analyses of matter effects on solar neutrino propagation. Precise
numerical values for $n_e$ are available at 
\label{fig:ne}}
\end{figure}

\begin{table}[!t]
\centering
\singlespace
\caption[]{\baselineskip=12pt
The electron number density versus radius for the standard
solar model.  The tabulated values are $\log (n_e/N_A)$, where $n_e$
is measured in number per ${\rm cm^3}$ and $N_A$ is Avogadro's
number. A more extensive numerical file of $n_e$, contain electron
number densities at $2493$ radial shells, is available at
http://www.sns.ias.edu/$\sim$jnb. \label{tab:ne}}
\begin{tabular}{ccccc}
\tableline\tableline
$R/R_\odot$&$\log (n_e/N_A)$&&$R/R_\odot$&$\log (n_e/N_A)$\\
\tableline
0.01& 2.008E+00&& 0.55& $-$1.527E$-$01\\
0.05& 1.956E+00&& 0.60& $-$3.605E$-$01\\
0.10& 1.827E+00&& 0.65& $-$5.585E$-$01\\
0.15& 1.662E+00&& 0.70& $-$7.428E$-$01\\
0.20& 1.468E+00&& 0.75& $-$9.098E$-$01\\
0.25& 1.249E+00&& 0.80& $-$1.099E+00\\
0.30& 1.012E+00&& 0.85& $-$1.330E+00\\
0.35& 7.687E$-$01&& 0.90& $-$1.642E+00\\
0.40& 5.269E$-$01&& 0.95& $-$2.164E+00\\
0.45& 2.914E$-$01&& 1.00& $-$6.806E+00\\
0.50& 6.466E$-$02\\
\tableline
\end{tabular}
\end{table}

Figure~\ref{fig:ne} and Table~\ref{tab:ne} give the electron number
density as a function of solar radius for the Standard solar model
(BP2000).
A much more extensive numerical file of the electron number density versus
radius is available at http://www.sns.ias.edu/$\sim$jnb; this file
contains the computed values of the electron number density at $2493$
radial shells.

We see from Figure~\ref{fig:ne} that for typical values of the neutrino
parameters that allow the so-called LMA and SMA MSW solutions which fit all of the currently
available solar neutrino data (e.g., Bahcall, Krastev, \& Smirnov
1998),\nocite{bahcall98} the experimentally most important $^8$B neutrinos ($E \geq
5~{\rm MeV}$) satisfy the resonance condition,
Eq.~(\ref{eq:mswresonance}), at radii that are smaller than the radius
of the convective zone. For the so-called LOW MSW solutions and for all MSW solutions with $\theta_V \sim \pi/4$,
the resonance radius falls in the outer part of the sun.

We have not previously published accurate values for the electron
density in the outer parts of the sun, $r ~ \geq~ 0.8 R_\odot$.
The straight line in Figure~\ref{fig:ne} is an approximation to the
electron number density in the standard solar model of Bahcall \&
Ulrich (1988)\nocite{bu88} (see Bahcall 1989).\nocite{bahcall89} 
This approximation, which has been
used for over a decade by  different groups analyzing solar
neutrino data, is

\begin{equation}
n_e/N_A ~=~ 245 \exp (-10.54 R/R_\odot) ~~{\rm cm^{-3}} .
\label{eq:nestraight}
\end{equation}
Figure~\ref{fig:ne} shows that
the approximation given in Eq.~(\ref{eq:nestraight}) fails badly in the
outer regions of the sun. Recently, several different analyses have
been published using the electron number density shown in
Figure~\ref{fig:ne} (or, more precisely, the computer file for $n_e(r)$
which is available at http://www.sns.ias.edu/$\sim$jnb ).

\subsection{The number density of scatterers of sterile neutrinos}
\label{subsec:steriledensity}

The effective density of particles for interacting with sterile (right
handed) neutrinos is not the electron number density discussed in the
previous subsection, but rather $n_{\rm sterile}$ (see Mikheyev \&
Smirnov 1986;\nocite{Mi86} Lim \&
Marciano 1988;\nocite{lim88} Barger et al. 1991)\nocite{barger91}, where

\begin{figure}[!b]
\centerline{\psfig{figure=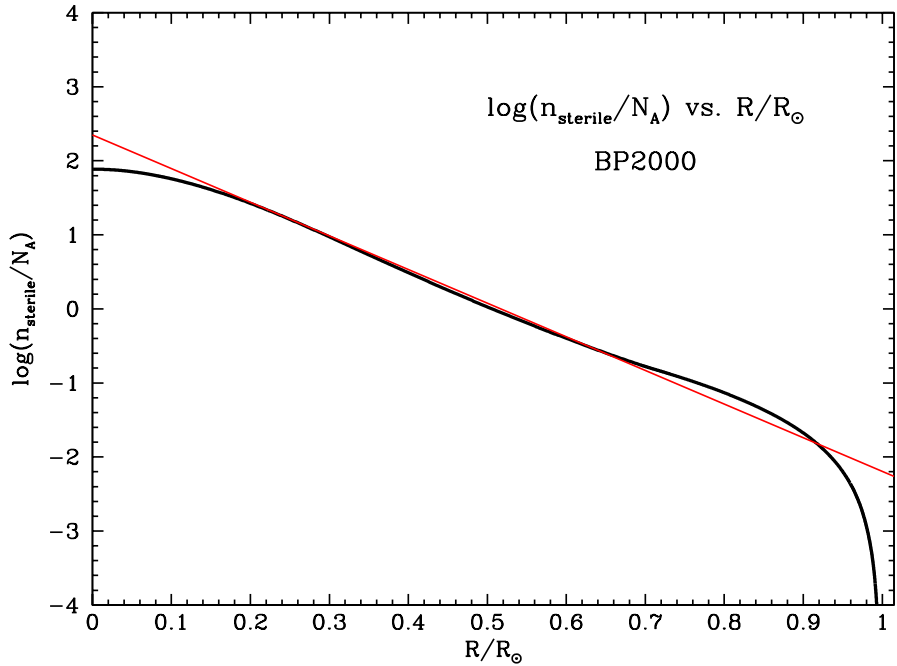,width=5in,angle=270}}
\caption{\baselineskip=12pt
The number density, $n_{\rm sterile}$, of scatterers of sterile
neutrinos  versus solar radius for
the Standard solar model (BP2000). The straight line 
 in Figure~\ref{fig:nsterile} is given by an equation of the same form
as  Eq. (\ref{eq:nestraight}) except that the coefficient for $n_{\rm sterile}/N_A$ is $223$ (instead of $245$ for 
$n_e/N_A$).}
\label{fig:nsterile}
\end{figure}

\begin{equation}
n_{\rm sterile} ~=~n_e ~-~ 0.5 \times n_{\rm neutrons},
\label{eq:steriledefn}
\end{equation}
and $n_{\rm neutrons}$ is the number density of neutrons. Since
nearly all of the neutrons in the sun are either in $^4$He or in
heavier elements with $Z \simeq A/2$, it is easy to derive an analytic
expression that relates $n_{\rm sterile}$ and $n_e$. One obtains

\begin{equation}
n_{\rm sterile} ~=~n_e\left(\frac{1 + 3X}{2(1 + X)}\right).
\label{eq:sterilerelation}
\end{equation}

Figure~\ref{fig:nsterile} and Table~\ref{tab:nsterile} give the radial
distribution of $n_{\rm sterile}$ in the Standard solar model. 
The functional form of $n_{\rm sterile}(r)$ is similar to the
function form of $n_e(r)$. The straight line in
Figure~\ref{fig:nsterile} is given by an equation of the same form as 
Eq. (\ref{eq:nestraight}) that describes $n_e(r)$ except that the
coefficient for $n_{\rm sterile}/N_A$ is $223$ (instead of $245$ for 
$n_e/N_A$).

\begin{table}[!t]
\centering
\singlespace
\caption[]{\baselineskip=12pt
The sterile  number density, $n_{\rm sterile}$, 
 versus radius for the standard
solar model.  The tabulated values are $\log (n_{\rm sterile}/N_A)$, 
where $n_{\rm sterile}$
is measured in number per ${\rm cm^3}$ and $N_A$ is Avogadro's
number. A more extensive numerical file of $n_{\rm sterile}$
 containing values of $n_{\rm sterile}$
at $2499$ radial shells, is available at
http://www.sns.ias.edu/$\sim$jnb. \label{tab:nsterile}}
\begin{tabular}{ccccc}
\tableline\tableline
$R/R_\odot$&$\log (n_e/N_A)$&&$R/R_\odot$&$\log (n_e/N_A)$\\
\tableline
 0.01& 1.885E+00&&   0.55& -1.901E-01\\
   0.05& 1.853E+00&&   0.60& -3.978E-01\\
   0.10& 1.757E+00&&   0.65& -5.956E-01\\
   0.15& 1.611E+00&&   0.70& -7.777E-01\\
   0.20& 1.425E+00&&   0.75& -9.436E-01\\
   0.25& 1.209E+00&&   0.80& -1.133E+00\\
   0.30& 9.731E-01&&   0.85& -1.364E+00\\
   0.35& 7.303E-01&&   0.90& -1.676E+00\\
   0.40& 4.887E-01&&   0.95& -2.198E+00\\
   0.45& 2.536E-01&&   1.00& -6.839E+00\\
   0.50& 2.701E-02\\
\tableline
\end{tabular}
\end{table}

The number density $n_{\rm sterile}$ is about $25$\% smaller than the
electron number density, $n_e$, in the center of the sun, where helium
is most abundant. In the central and outer regions of the sun, $n_{\rm
sterile}$ is about $9$\% less than $n_e$. Since the slopes of the
straight-line fitting functions are the same in Figure~\ref{fig:ne}
and Figure~\ref{fig:nsterile}, the neutrino survival probabilities are
the same for sterile and for active neutrinos as long as the adiabatic
approximation is valid (see, e.g., \S~9.2 of Bahcall 1989).\nocite{bahcall89}

\subsection{Neutrino fluxes as a function of solar age}
\label{subsec:fluxesvsage}

\begin{figure}[!t]
\centerline{\psfig{figure=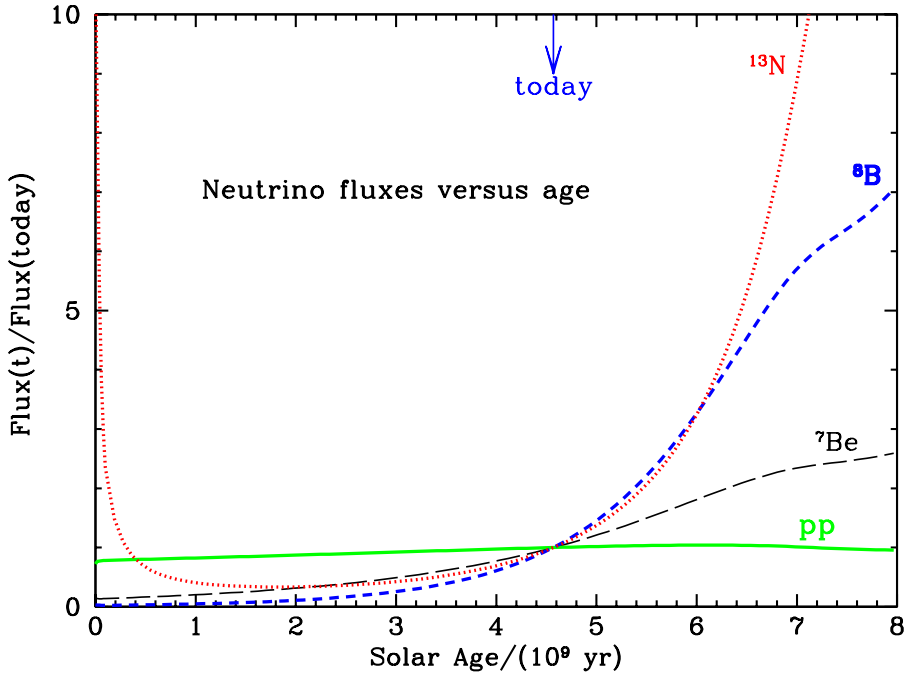,width=5in,angle=270}}
\caption{\baselineskip=12pt
The $pp$, $^7$Be, $^8$B, and $^{13}$N neutrino fluxes as a
function of solar age. The figure shows the Standard model
ratios of the fluxes divided by their values at $4.57 \times
10^9$ yr. The $pp$ flux is represented by a solid line, the $^7$Be
flux by a line of long dashes, the $^8$B flux by short dashes, and the
$^{13}$N flux by a dotted line.
\label{fig:fluxesvsage}}
\end{figure}

Figure~\ref{fig:fluxesvsage} shows the most important solar neutrino
fluxes, the $pp$, $^7$Be, $^8$B, and $^{13}$N fluxes, as a function of
solar age. The fluxes displayed in the figure were computed using the
Standard model and are normalized by dividing each flux by its value
at the present epoch, labeled by the word `today' in the figure
(cf. Fig.~10 of Guenther and Demarque 1991\nocite{demarque91} for a
similar figure plotted on a logarithmic scale).

The $pp$ flux is relatively constant over the entire $8$ billion years
show in Figure~\ref{fig:fluxesvsage}. At the beginning of its lifetime,
the $pp$ flux is about $75$\% of its current value and reaches $90$\%
of its present value after $2.6$ billion years. At the current epoch,
the flux is changing very slowly, about $4$\% per billion years. The
$pp$ flux reaches a maximum, $4$\% larger than its current value, at
a solar age of $6.0$ billion years and then declines slowly and
steadily to $96$\% of its present value at an age of  $8$ billion years.

The $^7$Be and $^8$B neutrino fluxes increase monotonically and by
larger amounts than the $pp$ flux. Both the $^7$Be and the $^8$B
fluxes begin with very low fluxes relative to their current values,
$14$\% and $3$\%, respectively, of their intensities at $4.57 \times
10^7$ yr. At a solar age of $8$ billion years, the $^7$Be neutrino flux
is $2.6$ times larger than it is today and the $^8$B neutrino flux is
$7.1$ times larger than today.  At the current epoch, the $^7$Be flux
is increasing by about $65$\% per billion years and the $^8$B flux is
increasing faster, about 120\% per
billion years.

The $^{13}$N neutrino flux has the most interesting time
dependence. In the first $10^8$ y on the main sequence, the $^{13}$N
flux is much larger than its current value because $^{12}$C has not
yet been burned to the equilibrium value appropriate for the CNO
cycle. The reaction $^{12}{\rm C}(p,\gamma)^{13}$N occurs relatively
often in this early stage of solar evolution and the neutrino flux
from $^{13}$N beta-decay has a peak value of about $11$ times its
current flux. The minimum $^{13}$N flux, $33$\% of its present value,
is attained at a solar age of $1.8$ billion years. Thereafter, the
$^{13}$N flux increases steadily as the central temperature of the
solar model increases and reaches an intensity of $18$ times its
current value at a solar age of $8$ billion years.

\section{Sound speeds}
\label{sec:soundspeeds}
Section~\ref{subsec:panoramic} presents a panoramic view of the
predicted Standard model sound speeds and compares the observations
and the calculations on a scale that is relevant for interpreting
solar neutrino experiments. In \S~\ref{subsec:zoom}, we compare on
a zoomed-in scale the Standard model calculations with the results of
six helioseismological measurements. The zoomed-in scale used in this
subsection can reveal fractional discrepancies between calculations
and observations that are smaller than $0.1$\%.  In
\S~\ref{subsec:rotation}, we consider a particular solar model that
includes rotation in a plausible way. This model smoothes the
composition discontinuity at the base of the convective zone, which
locally improves the agreement with the measured sound speeds but
worsens the overall rms agreement (within the uncertainties allowed by
input data).  For neutrino emission, the predictions of the
rotation model are not significantly different from the Standard
 model predictions.

Some authors (see, e.g., Guenther \& Demarque 1997\nocite{guen97})
compare their solar models directly with the $p$-mode oscillation
frequencies rather than with inverted quantities such as the sound
speed. The reader is referred to the Guenther and Demarque paper for a
discussion of the direct comparison method, its application, and
additional references.

We have chosen to use the sound speed profile because the inversion
process that produces the inferred sound speeds allows one to remove
the uncertainties, common to all $p$-mode oscillation frequencies,
that arise from the near-surface regions of the sun. These common
uncertainties are due to the treatment of convection, turbulence, and
non-adiabatic effects.  Inversion techniques are designed to minimize
the effects of these outer layers (see, e.g., Basu,
Christensen-Dalsgaard, Hernandez, and Thompson 1996)\nocite{basu96}
Moreover, the sound speed profile summarizes in a robust way the
results obtained for many thousands of oscillation frequencies.
Finally, the inversion procedure allows one to isolate different
regions of the sun, which is important in the context of discussions
regarding solar neutrinos. The neutrinos are produced deep in the
solar interior. In BPB2000, we have discussed in detail the systematic
uncertainties and assumptions related to the inversion for the sound
speeds and for the less-accurately determined density.

\begin{figure}[!b]
\centerline{\psfig{figure=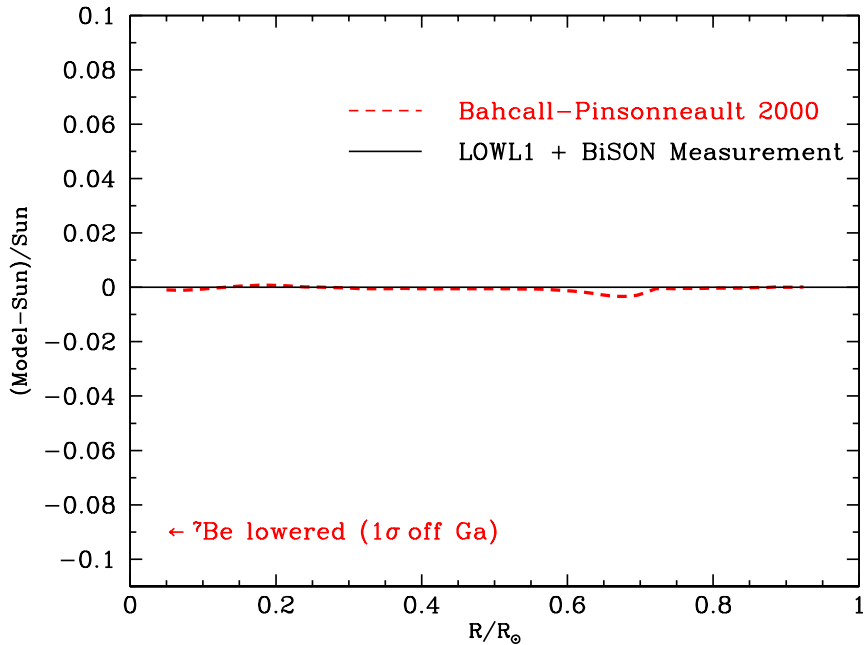,width=5in,angle=270}}
\caption[]{\baselineskip=12pt
Predicted versus measured sound speeds.  The figure shows
the excellent agreement between the calculated sound speeds for the
Standard solar model (BP2000) and the helioseismologically measured
(Sun) sound speeds.  The horizontal line at $0.0$ represents the
hypothetical case in which the calculated sound speeds and the
measured sound speeds agree exactly everywhere in the sun.  The rms
fractional difference between the calculated and the measured sound
speeds is $0.10$\% for all solar radii between between $0.05 R_\odot$
and $0.95 R_\odot$ and is $0.08$\% for the deep interior region, $r
\leq 0.25 R_\odot$, in which neutrinos are produced.
\label{fig:diffbp00best}}
\end{figure}

\subsection{Sounds speeds: panoramic view}
\label{subsec:panoramic}

Figure~\ref{fig:diffbp00best} shows the fractional differences between
the calculated sound speeds for the Standard model and what may be the
most accurate available sound speeds measured by helioseismology, the
LOWL1~+~BiSON Measurements presented in Basu et al. (1997).\nocite{LOWL1+BiSON}  These
sound speeds are derived from a combination of the data obtained by
the Birmingham Solar Oscillation Network (BiSON; cf. Chaplin et
al. 1996)\nocite{ch96} and the Low-$\ell$ instrument (LOWL; cf. Tomczyk et al. 1995a,b).\nocite{to95,tom95}

\begin{figure}[!b]
\centerline{\psfig{figure=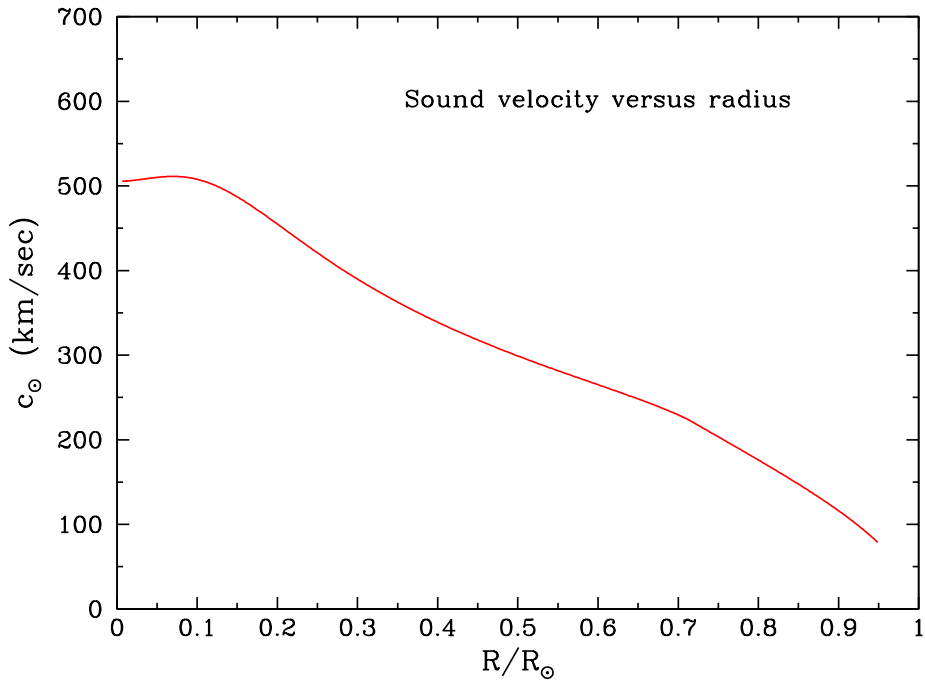,width=5in,angle=270}}
\caption[]{\baselineskip=12pt
The solar sound speed versus the solar radius.  The figure
shows the calculated solar sound speed versus radius for the standard
solar model, Bahcall-Pinsonneault (2000). To an accuracy of about
$0.5~{\rm km/s}$ the calculated and the observed sound speeds are the
same (see Figure ~\ref{fig:differences6} and Figure~\ref{fig:differences5}).
\label{fig:soundspeed}}
\end{figure}

The rms fractional difference between the calculated and the measured
sound speeds is $10.4 \times 10^{-4}$ over the entire region in which the
sound speeds are well measured, $0.05 R_\odot ~\leq~ r~ \leq~ 0.95
R_\odot$. In the solar core, $0.05 R_\odot ~\leq~ r~ \leq~ 0.25
R_\odot$ (in which about $95$\% of the solar energy and neutrino flux
are produced in a standard solar model), the rms fractional difference
between measured and calculated sound speeds is $6.3 \times 10^{-4}$.
The Standard model sound speeds agree with the measured sound speeds
to $0.1$\%  whether or not one limits the comparison to the solar
interior or averages over the entire sun. Systematic uncertainties
$\sim 3 \times 10^{-4}$ are contributed to the sound speed profile by
each of three sources:  the assumed reference model, the width of the
inversion kernel, and the measurement errors (see BPB2000). 

The vertical scale of Figure~\ref{fig:diffbp00best} was chosen so as to
include the arrow marked ``$ {\rm ~ ^7Be ~lowered~} (1 \sigma {\rm ~off~
Ga}$).'' This arrow indicates the typical difference between solar model
speeds and helioseismological measurements that would be expected if
the discrepancy between the gallium solar neutrino measurements and
the predictions in Table~\ref{tab:bestestimate} were due to errors in
the solar physics of the standard solar model (see discussion in
BP98).

Figure~\ref{fig:soundspeed} and Table~\ref{tab:soundspeeds} give the
sound speeds versus the solar radius that are calculated using the
Standard solar model. The sound speed declines from about $500$ km/s
in the solar core to about $100$ km/s at $0.95 R_\odot$. An extensive
table of the Standard model sound speeds is available at
http://www.sns.ias.edu/$\sim$jnb .

\begin{table}[!htb]
\centering
\singlespace
\caption[]{\baselineskip=12pt
\label{tab:soundspeeds} The sound speed as a function of radius
for the Standard solar model. The Sound speeds are given in units of
100 km/s.}
\begin{tabular}{ccccc}
\tableline\tableline
$R$&$c_s$&&$R$&$c_s$\\
$(R_\odot)$&(100 km/s)&&$(R_\odot)$&(100 km/s)\\
\tableline
0.01& 5.057&& 0.50& 2.990\\
0.05& 5.101&& 0.55& 2.817\\
0.10& 5.078&& 0.60& 2.651\\
0.15& 4.870&& 0.65& 2.483\\
0.20& 4.550&& 0.70& 2.295\\
0.25& 4.210&& 0.75& 2.035\\
0.30& 3.898&& 0.80& 1.761\\
0.35& 3.627&& 0.85& 1.476\\
0.40& 3.389&& 0.90& 1.162\\
0.45& 3.179&& 0.95& 0.778\\
\tableline
\end{tabular}
\end{table}

\subsection{Sounds speeds: zoom in}
\label{subsec:zoom}

\begin{figure}[!htb]
\centerline{\psfig{figure=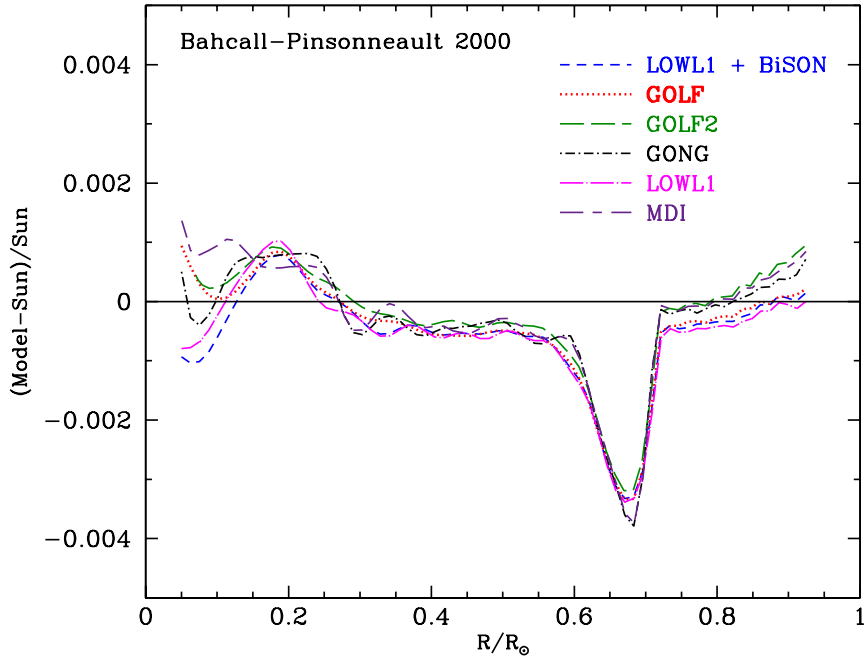,width=5in,angle=270}}
\caption[]{\baselineskip=12pt Six precise helioseismological
measurements versus BP2000.  The figure compares the fractional
difference between the sound speeds calculated for the Standard solar
model (BP2000) and the sound speeds in six helioseismological
experiments. The references to the helioseismological data are given
in the text. Systematic uncertainties due to the assumed reference
model and the width of the inversion kernel are each $\sim 0.0003$ (see
BPB2000).\label{fig:differences6}}
\end{figure}

Figure~\ref{fig:differences6} compares the results of six precise
observational determinations of the solar sound speed with the results
of our Standard solar model. The vertical scale has been expanded by a
factor of $21$ relative to Figure~\ref{fig:diffbp00best} in order to show
the small but robust discrepancies between the calculations and the
observations and to indicate the size of the differences between the
various measurements.  In the deep solar solar interior where
neutrinos are produced, $R/R_\odot \leq 0.25$, the differences between
the various observational determinations of the sound speed are
comparable to the differences between BP2000 and any one of the
measured sets of sound speeds.

The $p-$mode frequencies used in deriving the observed sound speeds
shown in Figure~\ref{fig:differences6} were obtained from a number of
different sources. In addition to the LOWL1~+~BiSON data described in
\S~\ref{subsec:panoramic}, we have used data from a number of other
sources. (1) Data from the first year of LOWL observations. The sound
speed inversions are described in Basu et
al. (1997)\nocite{LOWL1+BiSON} and are referred to as LOWL1 in this
paper. (2) Data from the Michelson Doppler Imager (MDI) instrument on
board the Solar and Heliospheric Observatory (SOHO) during the first
144 days of its operation (cf. Rhodes et al. 1997).\nocite{rho97} The
results of the sound speed inversions using these data are from Basu
(1998).\nocite{GONGMDI} (3) The frequencies obtained from the data
obtained by the Global Oscillation Network Group (GONG) between months
$4$--$14$ of its observations. The solar sound-speeds are from Basu
(1998).\nocite{GONGMDI} (4) Initial observation taken by the Global
Oscillations at Low Frequencies (GOLF) instrument on board SOHO,
combined with intermediate data from MDI.  We have labeled the sound
speeds obtained as GOLF1 and the solar sound speed results can be
found in Turck Chi\`eze et al. (1997).\nocite{GOLF} (5) More recent
data from GOLF (Thiery et al. 2000),\nocite{thierry2000} combined
with intermediate degree data obtained from the first 360 days
observations by the MDI instrument (Schou et al. 1997).\nocite{sch97}
The sound-speed results are described in Basu et
al. (2000).\nocite{basu00} These results have been labeled as GOLF2.

There are other helioseismological data sets that could have been
used. Of these, the data by Toutain et al. (1998)\nocite{toutain98} may be the most
relevant. This data set has been discussed extensively in Basu et
al. (2000).\nocite{basu00} Toutain et al. (1998)\nocite{toutain98} were the first to show the effect of
line-profile asymmetry on low-degree modes. The sound-speed inversions
in their paper were, however, very different from what had been
obtained previously. Basu et al. (2000)\nocite{basu00} showed that this difference
results primarily from the two modes, $l=2$, $n=6$ and $l=2$ $n=7$,
whose frequencies (and errors on the frequencies) were suspect. In a
later paper, Bertello et al. (2000)\nocite{bertello00} confirmed most of the low
frequency modes in the Toutain et al. (1998)\nocite{toutain98} data set, but not the two
questionable modes.  Additionally, they determined the frequencies of
some very low-frequency low-degree modes.  The sound-speed inversion
of Bertello et al. (2000)\nocite{bertello00} is similar to the inversion obtained in Basu
et al. (2000)\nocite{basu00} and in the present paper; the errors in the inversions
are also similar.  The similarity in the errors may appear somewhat
surprising at first glance given that the Bertello et al. frequencies
have very low errors, but the similarity in the overall uncertainties
is explained by the fact that the inversion errors are determined by
the error in the bulk of the frequencies (which are MDI
intermediate-$l$ frequencies), and not just the errors in the
low-degree modes.

The MDI and GONG sets
have good coverage of intermediate degree modes. The MDI set has
p-modes from $\ell = 0$ up to a degree of $\ell=194$ while the GONG
set has modes from $\ell = 0$ up to $\ell=150$.  However, both these
sets are somewhat deficient in low degree modes.  The LOWL1~+~BiSON
combination, on the other hand, has a better coverage of low degree
modes, but has modes from $\ell = 0$ only up to $\ell=99$. The GOLF data
sets only contain low-degree modes ($l=0,1,2$) and hence have to be combined
with other data before they can be used to determine the solar sound speed
profile.

\begin{figure}[!t]
\centerline{\psfig{figure=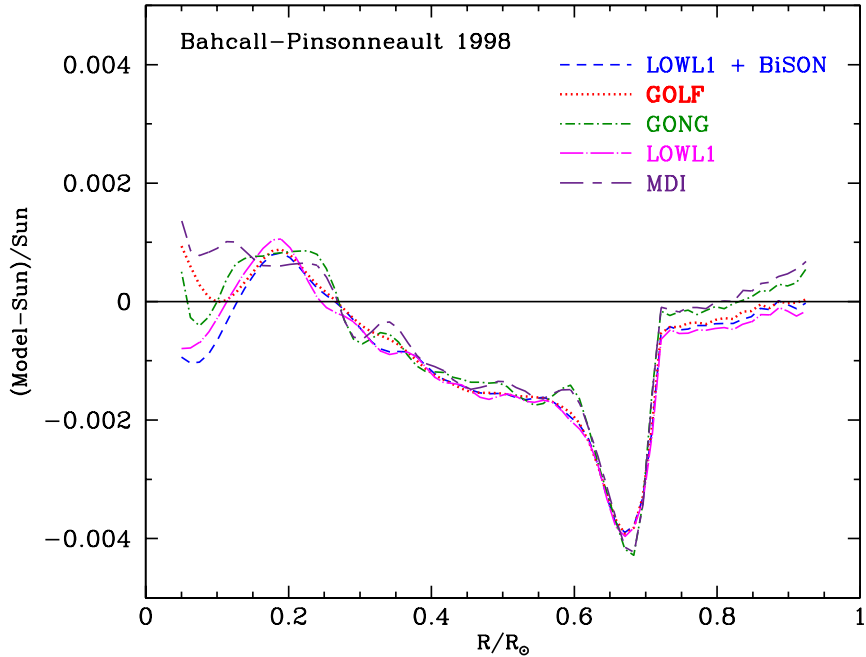,width=5in,angle=270}}
\caption[]{\baselineskip=12pt Five precise helioseismological
measurements versus BP98.  The figure compares the fractional
difference between the sound speeds calculated for the 1998 Standard
solar model (BP98)\nocite{ba98} and the sound speeds in five
helioseismological experiments. The references to the
helioseismological data are given in the text. (The GOLF2 data were
not available when this comparison was originally made.)  The rms
fractional difference between the calculated and the measured sound
speeds is $0.13$\% for all solar radii between between $0.05 R_\odot$
and $0.95 R_\odot$.
\label{fig:differences5}}
\end{figure}

Figure~\ref{fig:differences5} shows the somewhat less precise
agreement that was obtained between BP98 (Bahcall, Basu, \&
Pinsonneault 1998)\nocite{ba98} and the helioseismological data. The
BP98 model and the observed sound speeds agree in the solar interior
to about the same accuracy as for BP2000 and the observed
speeds. However, the BP98 model sound speeds are about $0.1$\% smaller
than the observed speeds in the broad intermediate region between $0.3
R/R_\odot$ and $0.7 R/R_\odot$ (cf. Figure~\ref{fig:differences6} and
Figure~\ref{fig:differences5} ).  Averaged over the entire region over
which good measurements are available, $0.05 R_\odot$ to $0.95
R_\odot$, the rms fractional difference between the BP98 model and the
LOWL1~+~BiSON sound speeds is $13 \times 10^{-4}$, which sound be
compared with a rms difference of $10.4 \times 10^{-4}$ for the BP2000
model (see \S~\ref{subsec:panoramic})\footnote{The Standard model
described in this paper differs from the BP98 model in two respects,
both discussed in \S~\ref{subsec:definitionstandard}, that are
significant for helioseismology: the correction of the opacity
interpolation error and the slightly different heavy element mixture
adopted for the BP2000 model. For helioseismology, the only
significant different between GN93 and BP98 is the correction of the
interpolation error; the heavy element abundances used for both models
are the same (with $Z/X = 0.0245$). When a figure like
Figure~\ref{fig:differences6} was constructed using GN93 instead of
BP2000, the general form of the small differences was practically the
same. In fact, the rms difference averaged over the whole sun between
GN93 and LOWL1 + BiSON is $0.00086$, which is slightly better than the
value of $0.00104$ for BP2000 versus LOWL1 + BiSON. One therefore
obtains the correct impression by comparing
Figure~\ref{fig:differences6} and Figure~\ref{fig:differences5}.}.

In BP98, we speculated that the broad feature of disagreement at the
$0.1$\% level in Figure~\ref{fig:differences5} might be due to a
combination of small errors in the adopted radiative opacities or in
the equations of state.  We investigated the implications for solar
neutrino fluxes of the possibility that the small, broad discrepancy
was due to an opacity error and concluded that if this were the case
then the corrected solar model would predict $^7$Be and $^8$B neutrino
fluxes that are about $5$\% larger than the fluxes predicted by BP98.

Indeed, the origin of the broad discrepancy is due to an error in
interpolating the radiative opacity near the edges of the opacity
tables, as explained in \S~\ref{subsec:definitionstandard}. The
effect on the neutrino fluxes is somewhat smaller than we had
estimated, an increase (relative to BP98) of $2$\% for $^7$Be
neutrinos and $3$\% for $^8$B (cf. the results for GN93 in 
Table~\ref{tab:bestestimate} and
Table~1 of BP98).

\subsection{Sounds speeds: rotation}
\label{subsec:rotation}

\begin{figure}[!t]
\centerline{\psfig{figure=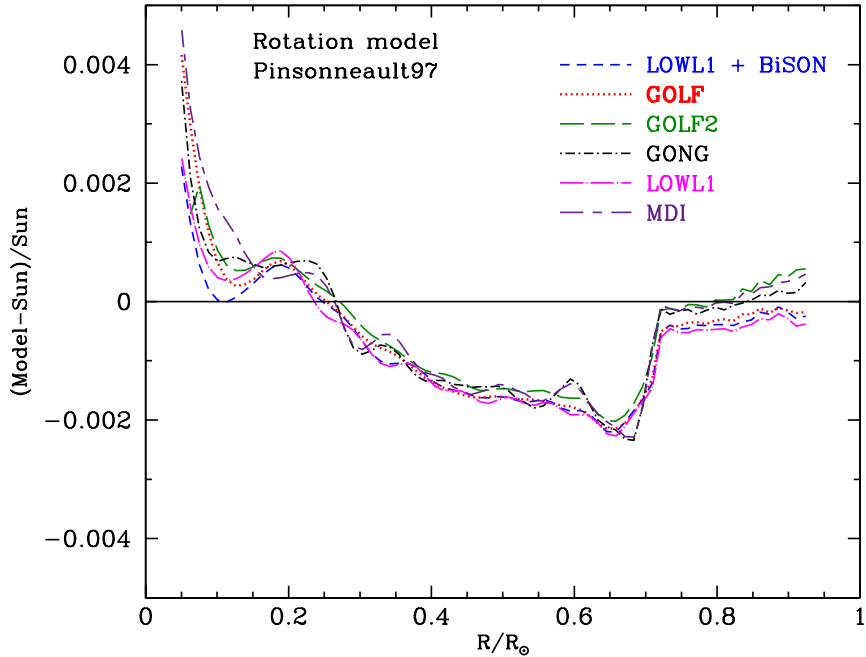,width=5in,angle=270}}
\caption[]{\baselineskip=12pt
Six precise helioseismological measurements versus the
Rotation model.
The figure compares the fractional difference between the sound speeds
calculated for the Rotational
solar model and the sound speeds in six helioseismological
experiments. The model was developed by Pinsonneault and collaborators
to explain the depletion of lithium. The references to the helioseismological data are given
in the text. \label{fig:differences6rot}}
\end{figure}

 Figure~\ref{fig:differences6rot} compares the sound speeds of the
 Rotation model with the six precise observed sets of sound speeds. In
 the region between $0.3 R_\odot$ and $0.6 R_\odot$,
 the agreement between the Rotation model and the helioseismological
 data is slightly less good than with the Standard solar model
 (compare Figure~\ref{fig:differences6} and
 Figure~\ref{fig:differences6rot}).  Quantitatively, over the entire region
 between $0.05 R_\odot$ to $0.95 R_\odot$, the rms fractional
 difference between the Rotation model and the LOWL1~+~BiSON sound
 speeds is $12 \times 10^{-4}$, which is to be compared with $10
 \times 10^{-4}$ for BP2000.  In the solar core ($\leq 0.25 R_\odot$),
 the two models, Rotation and BP2000, have almost identical rms
 fractional differences with respect to the LOWL1~+~BiSON sound speeds,
 $0.073$\% and $0.0064$\%, respectively.

\begin{figure}[!htb]
\centerline{\psfig{figure=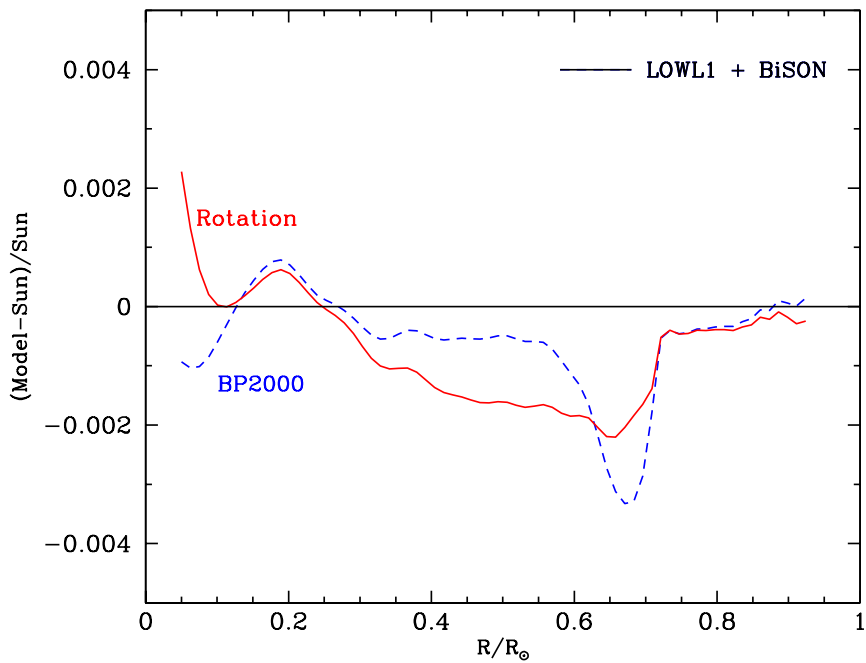,width=5in,angle=270}}
\caption[]{\baselineskip=12pt
BP2000 versus the Rotation model.  The figure compares the
fractional difference between the sound speeds determined from the
LOWL1~+~BiSON data with the sound speeds calculated for the BP2000 solar
model and the Rotational solar model. The BP2000 model agrees slightly
better with the measured sound speeds in the intermediate region
between $0.3 R_\odot$ and $0.6 R_\odot$.
\label{fig:differencesbprot}}
\end{figure}

 Figure~\ref{fig:differencesbprot} compares both the Rotation and the
 Standard model with just the LOWL1~+~BiSON sound speeds,  This figure shows
  that the Rotation model gives marginally better agreement
 with the measured sound speeds right at the base of the convective
 zone, comparable agreement in the deep interior, $r \leq 0.25
 R_\odot$, and slightly less good agreement in the intermediate region
 between $0.3 R_\odot$ and $0.6 R_\odot$.

The neutrino fluxes calculated
 with the Rotation model lie well within the estimated errors in the
 Standard model fluxes, as can be seen easily by comparing the fluxes
 and the errors given in Table~\ref{tab:bestestimate} and
 Table~\ref{tab:fluxesnine}. The $^7$Be flux for the Rotation model is
 $3$\% less than the Standard model $^7$Be and the $^8$B flux is $5$\%
 lower than the corresponding Standard model value. The Rotation model
 predicts a capture rate by ${\rm ^{37}Cl}$ that is 0.3 SNU less than
 the Standard model rate and a ${\rm ^{71}Ga}$ capture rate that is
 2 SNU less than the Standard model rate.

\section{Helium abundance and depth of the convection zone}
\label{sec:hepluscz}

Table~\ref{tab:yandcz} gives the calculated present-epoch values for
the depth of the convective zone,
R(CZ), and the surface helium abundance, $Y_s$, for all $11$ of
the solar models considered in this
paper.

\begin{table}[!t]
\centering
\singlespace
\caption[]{\baselineskip=12pt
Helium abundance and the depth of the convective zone.  The
calculated present-day helium abundance on the surface and the
convective zone depth are given for 11 solar models discussed in this
paper.  The quoted errors in the measured values of $Y$ and $R(CZ)$
represent best estimates of the systematic uncertainties but cannot be
interpreted rigorously in terms of $1\sigma$ or $3\sigma$
errors.\label{tab:yandcz}}
\begin{tabular}{lcc}
\tableline\tableline
\multicolumn{1}{c}{Standard}&$Y_s$&$R(CZ)$\\
\tableline
Standard&0.244&0.714\\
NACRE&0.244&0.713\\
AS00&0.239&0.714\\
GN93&0.245&0.712\\
Pre-MS&0.246&0.713\\
Rotation&0.248&0.714\\
${\rm Radius_{-78}}$&0.245&0.712\\
${\rm Radium_{-508}}$&0.245&0.712\\
No diffusion&0.266&0.726\\
Old physics&0.248&0.712\\
$S_{34} = 0$&0.242&0.715\\
Mixed&0.254&0.732\\
\noalign{\medskip}
Measured&$0.249\pm 0.003$&$0.713\pm 0.001$\\
\tableline
\end{tabular}
\end{table}

The observed values determined using measurements of $p$-modes are
$R(CZ) = (0.713 \pm 0.001) R_\odot$ (Basu \& Anita
1995;\nocite{basu95} for earlier work see Christensen-Dalsgaard,
Gough, \& Thompson 1991)\nocite{jcd91} and $Y_s = 0.249 \pm 0.003$
(Basu \& Anita 1997;\nocite{basu97} see also Richard et
al. 1996).\nocite{ri96}

The quoted errors are systematic; the statistical uncertainties are
much smaller. For example, the cited uncertainty for $Y_s$ is designed
to span the two values obtained when using two different equations of
state (cf. Basu \& Anita 1997).\nocite{basu97} There is no rigorous
way of establishing a confidence level based upon agreement within,
for example, one or two times the estimated systematic uncertainty. We
point out that the recent reassessment of $Z/X$ by Grevesse and Sauval
(1998)\nocite{gre98} resulted in a value of $Z/X = 0.0230$ which
differs from the helioseismologically recommended value of $Z/X =
0.0245 \pm 0.0008$ by twice the quoted uncertainty of the
helioseismological determination.

As a rule-of-thumb, we shall regard
agreement within two times the quoted systematic uncertainty as
`satisfactory' and agreement within one times the quoted systematic
uncertainty as `excellent'.

Nine of the eleven solar models considered in this paper give
excellent or satisfactory agreement with the observed depth of the
convective zone. The only exceptions to the good agreement with the
measured convective zone depth are the No Diffusion and the Mixed
models, which are both strongly disfavored by the helioseismological
measurements.  

The No Diffusion, $S_{34} = 0$, and AS00 models are the only ones that
are not within twice the quoted uncertainty in the measured surface
helium abundance. 
The No Diffusion model yields a surface helium abundance that is more
than five times the quoted uncertainty away from the
helioseismological measurement, which very strongly disfavors the No
Diffusion model.
The surface helium abundance calculated for the AS00
model is also rather far from the helioseismological value, more than three
times the quoted uncertainty in the helioseismological
determination. This discrepancy should be examined more fully in
future years as the abundance determinations are refined and the
helioseismological determination of the helium abundance is repeated
and the input data to the helioseismological analysis is varied over
a wide range of allowed possibilities.

\section{Discussion and summary}
\label{sec:discussion}

This paper provides new information about four topics: 1) the
characteristics of the Standard solar model at the current epoch (see
\S~\ref{subsec:discusscurrent} below); 2) time dependences of important
characteristics of the Standard solar model
(\S~\ref{subsec:discusstime}); 3) neutrino fluxes and related
quantities for standard and variant solar models
(\S~\ref{subsec:neutrinocharacteristics}); and 4) measured versus
calculated solar sound speeds (\S~\ref{subsec:discusssoundspeeds}).
Extensive numerical data that are useful for applications are
available at http://www.sns.ias.edu/$\sim$jnb .

Just for fun, we provide our favorite list, our ``top three,'' among the
many disparate results presented in this paper. Our top three results
are listed in
\S~\ref{subsec:fun}.

\subsection{Standard solar model: current epoch}
\label{subsec:discusscurrent}
We present detailed numerical tabulations of the computed
characteristics of our Standard solar model, which is defined and
discussed in \S~\ref{sec:standard}.  These tables include, as a
function of the solar radius, the enclosed mass fraction, the
temperature, mass density, electron number density, the pressure, and
the luminosity fraction created in a given spherical shell, as well
as the mass fractions of $^1$H, $^3$He, $^4$He, $^7$Be, $^{12}$C,
$^{14}$N, and $^{16}$O. Over the years, previous numerical versions of
our Standard model have been used for a variety of purposes that range
from comparisons with other stellar evolution codes, estimating the
importance in the sun of newly considered physical effects, searching
for possible instabilities in the sun, comparison with
helioseismological measurements, and the calculation of processes
(especially the MSW effect) that
influence the propagation of solar neutrinos.

In the past, we have published in hard copy form increasingly more
detailed and precise numerical tables of the characteristics of the
solar interior. The capabilities of current calculations and the
requirements of some of the most interesting applications have made
complete hard copy publication no longer appropriate. We have therefore limited
ourselves in \S~\ref{sec:standard} to describing briefly the
ingredients we use in calculating the current Standard model.
We present the numerical results in exportable data files that are
available at http://www.sns.ias.edu/$\sim$jnb .

\subsection{Standard solar model: time dependences}
\label{subsec:discusstime}

For the first time in this series of papers, we have focused,
especially in \S~\ref{sec:timedependences}, on details of the time
dependence of important characteristics of the Standard solar model.

The total luminosity in the Standard model increases by $48$\% from
the zero-age main sequence stage to the present epoch. Over the same
period, the effective temperature varies by only $\pm 1.3$\%.  These
predictions constitute constraints on models for the evolution of the
earth.

The predicted time evolution of the solar luminosity is
robust. Figure~\ref{fig:lumnormalized} shows that all solar models,
even those models with deficient physics that are strongly disfavored
by helioseismological measurements, predict essentially the same
luminosity evolution. The average rms deviation of the deviant models,
the Mixed, No diffusion, and $S_{34} = 0$ models, from the standard
solar model luminosity is only $1$\% over the history of the sun from
$1$ Gyr to the current epoch (see \S~\ref{subsec:radiusetal} for
more details).
 
Table~\ref{tab:separations} presents the calculated large and small
separations of the $p$-mode frequencies as a function of age for the
standard solar model.

We have presented in \S~\ref{sec:timedependences} the time
evolution of some of the principal physical quantities characterizing
the solar core (the central temperature, density, pressure, and
hydrogen mass fraction, as well as the fractions of the solar
luminosity generated by different nuclear reactions). We also present
 the evolution of important quantities at the base of
the convective zone (radiative opacity, temperature, density, and
pressure).  We hope that these data and the scaling
relations we have inferred will be sufficient to permit a future
physical understanding of the time dependences using analytic and
semi-analytic arguments.

We find some simple scaling relations. For example, the solar
luminosity, $L_\odot(t)$, is approximately related to the solar
radius, $R_\odot(t)$, as $L_\odot(t) \propto R_\odot(t)^{2.5}$. The
depth of the convective zone, $R({\rm CZ},t)$, scales as $R({\rm
CZ},t) \propto R_\odot(t)$, and the central temperature, $T_c(t)$,
shows a similar behavior, $T_c(t) \propto R_\odot(t)$.  Moreover, we
find that the mass of the convective zone, $M({\rm CZ},t)$ satisfies
$M({\rm CZ},t) \propto R_\odot(t)^{-2}$.  The effective temperature is
approximately constant, varying by only $\pm 0.7\%$ from a solar age
of 2 billion years to 8 billion years.

These results make predictions that are potentially testable.  In
principle, the measurement of the luminosity (by astrometry) of a star
with the same mass and chemical composition ($Z/X$) as the sun would
allow the prediction of the star's effective temperature,   the depth
and mass of the convective zone, and the large and small separations
of the $p$-mode frequencies 
(see for example Monteiro, Christensen-Dalsgaard,  \& Thompson
2000\nocite{mont00}  
and Christensen-Dalsgaard 1997\nocite{jcd97}).
In practice, it is difficult to make
measurements sufficiently accurately to make possible precise tests of
stellar evolution theory.  For an appraisal of both the potential and
the difficulty of making such measurements, the reader is referred to
the recent papers by Guenther and Demarque (2000)\nocite{guen00} 
and Morel, Provost, Lebreton, Th\'evenin, \& Berthomieu
(2000)\nocite{morel2000} on the binary pair of approximately solar
mass stars, $\alpha$ Centauri AB ($1.1 M_\odot$ and $0.9 M_\odot$).

\subsection{Neutrino fluxes and related quantities}
\label{subsec:neutrinocharacteristics}

Figure~\ref{fig:ne} and Table~\ref{tab:ne} give the electron number
density as a function of position in the sun for the Standard solar
model.  The distribution of the electron density is required to
compute the probability for matter-induced oscillations between active
neutrinos.  Similarly, Figure~\ref{fig:nsterile} and
Table~\ref{tab:nsterile} give the radial distribution of the number
density of scatterers of sterile neutrinos, $n_{\rm sterile}$, in the
Standard solar model.  We have not previously published precise values
for the electron number density, or of $n_{\rm sterile}$, in the outer
regions of the sun. The outer regions are relevant for large mixing
angle neutrino oscillations with relatively low neutrino mass
differences ($\Delta m^2 < 10^{-8} {\rm eV^2}$).

Table~\ref{tab:bestestimate} presents the neutrino fluxes and the event
rates in the chlorine, gallium, lithium, and electron-scattering neutrino
experiments that are predicted by the Standard solar model. These
predictions assume that nothing happens to solar neutrinos after they
are produced.  The table also gives estimates of the uncertainties
in the fluxes and the event rates; \S~\ref{subsec:fluxestoday}
contains a discussion of the physical origin of the uncertainties, as
well as the software used to calculate the asymmetric error estimates.

How do the predictions of solar neutrino event rates compare with
experiment? Table~\ref{tab:numeasurements} compares the predictions of
BP2000 with the results of the chlorine, GALLEX + GNO, SAGE,
Kamiokande, and Super-Kamiokande solar neutrino experiments. This
table assumes nothing happens to the neutrinos after they are created
in the sun. The standard predictions differ from the observed rates by
many standard deviations. Because of an accidental cancellation, the
predicted solar neutrino event rates for BP2000 and BP98 are almost
identical (see \S~\ref{subsubsec:standardneutrino}).

Table~\ref{tab:fluxesnine} compares the neutrino fluxes and the
experimental event rates for all nine of the solar models whose
helioseismological properties were investigated in BPB2000, plus two
additional standard-like models considered here which have somewhat
different heavy-element to hydrogen ratios.  The seven standard-like
models (the first seven models in Table~\ref{tab:fluxesnine}) all
produce essentially the same neutrino predictions; the spreads in the
predicted $pp$, $^7$Be, and $^8$B fluxes are $\pm 0.7$\%. $\pm
3$\%. and $\pm 6.5$\% , respectively.  The calculated rates for the
seven standard-like solar models have a range of $\pm 0.45$ SNU for
the chlorine experiment and $\pm 2$ SNU for the gallium experiments.

The estimated total errors from external sources (see
Table~\ref{tab:bestestimate}), such as nuclear cross section
measurements and heavy element abundances, are about a factor of three
larger than the uncertainties resulting from the solar model
calculations.

We have investigated one possible source of systematic errors, the
relative weights assigned to different determinations of nuclear
fusion cross sections. We calculated the neutrino fluxes and predicted
event rates using the NACRE (Angulo et al. 1999)\nocite{angulo99}
fusion cross sections rather than the Adelberger et
al. (1998)\nocite{adel98} cross sections. The NACRE parameters lead to
slightly higher predicted event rates in solar neutrino
experiments. However, all changes in the neutrino fluxes and event
rates between the NACRE-based predictions and the Standard predictions
(based upon Adelberger et al. nuclear parameters) are much less than
the $1\sigma$ uncertainties quoted for the Standard model (cf.
Table~\ref{tab:bestestimate} and Table~\ref{tab:nacre}).

The neutrino event rates predicted by all seven of the standard-like
solar models considered here are inconsistent at the
$5 \sigma$ level (combined theoretical and experimental errors)
 with the results of the two gallium experiments, GALLEX
and SAGE, assuming no new physics is occurring. The inconsistency
with the chlorine experiment is similar but more complex to specify,
since the largest part of the theoretical uncertainty in the
calculated standard capture rate is due to the electron-type neutrinos
from $^8$B beta-decay.  The fractional uncertainty in the $^8$B flux
depends upon the magnitude of the flux created in the solar
interior. Moreover, the amount by which this $^8$B flux is reduced
depends upon the adopted particle physics scenario.

A similar level of inconsistency persists even for the ad hoc
deficient models, such as the $S_{34} = 0$ and Mixed models, that were
specially concocted to minimize the discrepancy with the neutrino
measurements.  For example, the calculated rates for the Mixed model
are $6.15^{+1.0}_{-0.85}$ SNU for the chlorine experiment and
$115^{+6.8}_{-5.1}$ SNU for gallium experiments. The Mixed model is
$4.2\sigma$ below the measured chlorine rate and $6.3\sigma$ below the
measured gallium rate. In addition, the deficient models are strongly
disfavored by the helioseismology measurements.

Figure~\ref{fig:fluxesvsage} shows the calculated time dependence of
the $pp$, $^7$Be, $^8$B, and $^{13}$N solar neutrino fluxes. At the
current epoch, the $pp$ flux is increasing at a rate of $4$\% per
billion years, and the $^7$Be, $^8$B, and $^{13}$N fluxes at the rates
of $45$\%, $90$\%, and $99$\% per billion years, respectively. Since
the age of the sun is estimated to be uncertain by only
$5\times10^{-3}$ billion years (see the Appendix by Wasserburg in
Bahcall \& Pinsonneault 1995),\nocite{bp95} the age of the sun does not represent
a significant uncertainty for solar neutrino predictions.

\subsection{Sound speeds}
\label{subsec:discusssoundspeeds}

Figure~\ref{fig:diffbp00best} shows the excellent agreement between the
helioseismologically determined sound speeds and the speeds that are
calculated for the Standard solar model. The scale on this figure was
chosen so as to highlight the contrast between the excellent agreement
found with the Standard model and the two orders of magnitude larger
rms difference for a solar model
that could reduce significantly the solar neutrino problems. One would
expect characteristically a $9$\% rms difference between the
observations and the predictions of solar models that significantly
reduce the conflicts between solar model measurements and solar model
predictions. Averaged over the entire sun, the rms fractional difference is
only $0.10$\% between the Standard solar model sound speeds and the
helioseismologically-determined  sound speeds.  The agreement is even
better, $0.06\%$, in the interior region in which the luminosity and
the neutrinos are produced. 
Table~\ref{tab:soundspeeds} presents numerical values for
the sound speeds predicted by the Standard solar model at
representative radial positions in the sun.

All eight of the standard-like solar models considered in this paper
give acceptable agreement with the measured depth of the convective zone and
the surface helium abundance (see Table~\ref{tab:yandcz}).  Of the
twelve Standard, standard-like,  variant, and
deviant solar models considered here, only the No Diffusion and Mixed
models disagree strongly with the measured convective zone depth.

There are small but robust discrepancies between the measured and the
calculated solar sound speeds. Figure~\ref{fig:differences6} shows the
fractional differences between the Standard model sound speeds and the
speeds measured in each of six different determinations (using
$p$-mode data from LOWL1, BiSON, GONG, MDI, GOLF, and GOLF2). The
vertical scale for Figure~\ref{fig:differences6} is expanded $21$ times
compared to the vertical scale of  Figure~\ref{fig:diffbp00best}.   As can
be seen from Figure~\ref{fig:differences6}, the differences between
observed and measured sound speeds are comparable over much of the sun
to the differences between different measurements of the sound speed,
but there is a clear discrepancy near the base of the convective zone
that is independent of which observational data set is used.  There
may also be a less prominent discrepancy near $0.2 R_\odot$.

The agreement between the BP2000 sound speeds and the measured values
is improved over what was found earlier with the BP98
model. This improvement may be seen by comparing
Figure~\ref{fig:differences6} and Figure~\ref{fig:differences5}. The
improvement is striking in the region between $0.3R_\odot$ and
$0.7R_\odot$ and is due to the correction of an error in the interpolation
algorithm for the radiative opacity (cf. discussion in
\S~\ref{subsec:definitionstandard}).  

The Rotation model includes a prescription for element mixing that was
designed to explain the depletion of lithium. The calculated
differences between the Rotation and the Standard models represent a
reasonable upper limit to the effects that rotation, sufficient to
explain lithium depletion, might produce. The overall agreement
between the sound speeds of the Rotation model and the
helioseismologically determined sound speeds is slightly worse,
$0.12$\% rather than $0.10$\%, than the agreement obtained with the
Standard solar model. The differences between the solar neutrino
fluxes predicted between the Rotation model and the Standard model is
typically about $0.3\sigma_\nu$, 
where $\sigma_\nu$ represents the uncertainties
given in Table~\ref{tab:bestestimate} for the Standard model neutrino
fluxes.  We conclude that further improvements of the theoretical
calculations motivated by refinements of $p-$ mode oscillation
measurements are unlikely to significantly affect the calculated solar
neutrino fluxes.

\subsection{Top three results}
\label{subsec:fun}

Here are our favorite three results in this paper.

$\bullet$ The robust luminosity evolution of the sun (see
Figure~\ref{fig:lumnormalized}).

$\bullet$ The excellent agreement of the Standard model sound speeds
with the measured sound speeds on the scale relevant for solar
neutrino discussions (see Figure~\ref{fig:diffbp00best}).

$\bullet$ The simple relations as a function of time between the solar
radius, the solar luminosity, the depth of the convective core, and
the mass of the convective core (see
Eq.~\ref{eqn:lrproportional}--Eq.~\ref{eqn:tcoverr}).

\acknowledgments
We are grateful to many colleagues in solar physics, nuclear physics,
and particle physics for valuable discussions, advice, criticism, and
stimulation.  JNB is supported in part by an NSF grant \#PHY-0070928.

\end{document}